\documentclass[preprint,12pt,authoryear]{elsarticle}
\usepackage{amssymb}
\usepackage[utf8]{inputenc}   
\graphicspath{ {./figures/} }
\usepackage{hyperref}
\usepackage{floatrow}
\usepackage{verbatim} 
\usepackage{amsfonts}
\usepackage{booktabs}
\usepackage{multirow}
\usepackage{threeparttable}
\usepackage{tabularx}
\usepackage{makecell}
\usepackage{hyperref}
\usepackage{lineno}
\restylefloat{figure}
\restylefloat{table}

\journal{Expert Systems with Applications}

\begin{document}

\floatsetup[table]{capposition=top}

\begin{frontmatter}

\title{QUSL: Quantum Unsupervised Image Similarity Learning with Enhanced Performance}

\author[lable1]{Lian-Hui Yu}
\author[lable2]{Xiao-Yu Li$^\dagger$}
\author[lable3]{Geng Chen}
\author[lable1]{Qin-Sheng Zhu}
\author[lable4]{Hui Li}
\author[lable3,lable5]{Guo-Wu Yang}

\affiliation[lable1]{organization={School of Physics, University of Electronic Science and Technology of China},
            city={Cheng Du},
            postcode={610054}, 
            state={Si Chuan},
            country={China}}
\affiliation[lable2]{organization={School of Information and Software Engineering, University of Electronic Science and Technology of China},
	city={Cheng Du},
	postcode={610054}, 
	state={Si Chuan},
	country={China}}
\affiliation[lable3]{organization={School of Computer Science and Engineering, University of Electronic Science and Technology of China},
	city={Cheng Du},
	postcode={610054}, 
	state={Si Chuan},
	country={China}}
\affiliation[lable4]{organization={JILA and Department of Physics, University of  Colorado},
	city={Boulder},
	postcode={80309-0440}, 
	state={Colorado},
	country={USA}}
\affiliation[lable5]{organization={Advanced Cryptography and System Security Key Laboratory of Sichuan Province},
	city={Cheng Du},
	postcode={610103}, 
	state={Si Chuan},
	country={China}}

\begin{abstract}
Leveraging quantum properties to enhance complex learning tasks has been proven feasible, with excellent recent achievements in the field of unsupervised learning. However, current quantum schemes neglect adaptive adjustments for unsupervised task scenarios. This work proposes a novel quantum unsupervised similarity learning method — QUSL. Firstly, QUSL uses similarity triplets for unsupervised learning, generating positive samples by perturbing anchor images, achieving a learning process independent of classical algorithms. Subsequently, combining the feature interweaving of triplets, QUSL employs metaheuristic algorithms to systematically explore high- performance mapping processes, obtaining quantum circuit architectures more suitable for unsupervised image similarity tasks. Ultimately, QUSL realizes feature learning with lower quantum resource costs. Comprehensive numerical simulations and experiments on quantum computers demonstrate that QUSL outperforms state-of-the-art quantum methods. QUSL achieves over $50\%$ reduction in critical quantum resource utilization. QUSL improves similarity detection correlation by up to $19.5\%$ across multiple datasets, exhibiting robustness in NISQ environments. While using fewer quantum resources, QUSL shows potential for large-scale unsupervised tasks.
\end{abstract}

\begin{keyword}
Quantum machine learning \sep Image similarity detection \sep Metaheuristic algorithms \sep Unsupervised learning
\end{keyword}

\end{frontmatter}

\section{Introduction}
\label{introduction}

Quantum computing represents a paradigm shift in computational capabilities and has entered a stage known as Noisy Intermediate-Scale Quantum (NISQ) computing \cite{Knill, Preskill}. Quantum resources possess invaluable potential, and quantum computing demonstrates more apparent advantages when dealing with complex tasks \cite{Arute, HuangHYQuantumadvantage, HuangB, li2024new}. Quantum Machine Learning (QML) \cite{Biamonte} has emerged as a noteworthy research focus in the field of image processing, giving rise to representative models such as quantum neural networks (QNN) \cite{SchuldM, Schuld, CongI}. QML boasts several advantages \cite{SchuldM, XuY}, one of which is its ability to utilize a larger feature space, providing enhanced capabilities for modeling complex patterns in image processing.

One particularly challenging task in the field of image processing is image similarity detection, which has far-reaching implications for complex applications such as target tracking and object recognition in real-world scenarios \cite{WangX, ChengG}. It can be conceptualized as instances of image multiclassification problems characterized by a plethora of unknown classes, each containing only one member. Classical approaches encounter prominent challenges. With increasing task scale, the complexity and computational demands of image similarity detection escalate rapidly. This limits the feasibility of employing simple and intuitive algorithms \cite{ZhangL, PalubinskasG} in large-scale tasks, while the use of data dimensionality reduction techniques inevitably leads to information loss. This indirectly affects the generalization ability of unsupervised and self-supervised learning schemes \cite{BaiY, WangJ}, resulting in unacceptable learning costs for large-scale, high-resolution image datasets.

Due to the potential of quantum computing in image processing tasks, the first wave of quantum similarity algorithms \cite{DangY, LiuXA} and quantum machine learning models \cite{ZhouRG, YanF} specifically tailored to address challenges in image similarity analysis have been developed. Series works have focused on establishing unsupervised learning processes by referencing classical models. Jaderberg et al. \cite{jaderberg2022quantum} adopted SimCLR's unsupervised learning strategy \cite{chen2020simple}, enhancing the original network using QNNs. Silver et al.'s research \cite{SilverDSLIQ} introduced the first quantum unsupervised image similarity model, employing triplets\cite{ma2020fine} from contrastive learning to encode image pairs and likewise utilizing QNNs for feature extraction. Despite demonstrating promising performance, these works overlooked the necessity of adapting quantum learning circuits to suit specific task requirements. The critical challenge in employing quantum computing for image similarity tasks lies in the design and training of high-performance quantum circuits that are tailored to a specific task scenarios.

The design of quantum circuits for learning scenarios presents a specialized optimization challenge \cite{WangH, MohseniN}, aiming to minimize the usage of entanglement gates within the circuit structure's shallowest possible depth while achieving superior expressive power\cite{sim2019expressibility}. Under such task scenarios, obtaining gradient information and ensuring differentiability are challenging, at times entirely unknown for quantum circuits involving intricate quantum state evolution processes. Consequently, the introduction of metaheuristic algorithms with self-organization and self-learning characteristics has become an efficient approach for obtaining more advanced quantum circuits \cite{ZhangA, DingL, KrylovG, RasconiR}.

In light of the aforementioned challenges and limitations, our research proposes a novel quantum unsupervised similarity learning framework — QUSL. The core motivation of QUSL is to fully harness the high expressive power of quantum circuits to more effectively process unsupervised image similarity tasks, breaking free from the reliance on classical image similarity algorithms and the performance constraints imposed by parameterized quantum circuit templates.

QUSL introduce an evolutionary algorithm-driven method to explore quantum circuit architectures customized for dataset features. This approach enables the automatic design of quantum circuits that are more efficient than fixed-template parameterized quantum circuits, effectively capturing the intricate patterns within image datasets and enhancing the performance of image similarity detection.  Subsequently, we propose a perturbation-based strategy for constructing similarity triplets, allowing QUSL to independently learn and detect image similarities without relying on classical algorithms, fully leveraging the advantages of quantum computing. QUSL provides a more generalizable and scalable solution for image similarity detection.

The highlights of our research can be summarized as follows:

\begin{enumerate}
	\item QUSL harnesses an evolutionary algorithm-driven method to explore quantum circuit architectures customized for dataset features, achieving more efficient quantum image feature extraction with reduced circuit complexity and learning costs, demonstrating cross-scenario transferability. The \href{https://github.com/QUSL0414/QUSL/tree/main}{code} for QUSL is available. 
	\item Leveraging quantum circuit properties, QUSL incorporates perturbed images to independently construct quantum $A, P, N$-triplets for unsupervised image similarity detection, offering an adjustable similarity threshold without relying on classical algorithms.
	\item Numerical simulations and experiments on quantum computers across five datasets demonstrate QUSL's superiority over state-of-the-art quantum methods, reducing critical quantum resource utilization by over $50\%$ while enhancing similarity detection correlation up by $19.5\%$, showcasing its performance advantage in near-realistic scenarios.
\end{enumerate}

\section{Background}
\label{Background}

\subsection{Qubits and Quantum Circuits}
\label{Qubits and Quantum Circuits in NISQ}

At the core of quantum computing lies the quantum bit (qubit), which serves as the fundamental unit of information in quantum computation. Unlike classical bits in binary states $0$ and $1$, A quantum bit $|\psi\rangle$ can simultaneously exist in multiple states due to the principle of superposition, as depicted below
\begin{equation}
	|\psi\rangle=\alpha|0\rangle+\beta|1\rangle\quad s.t\ \alpha,\beta\in\mathbb{C}\ \&\ ||\alpha||^2+||\beta||^2=1.
\end{equation}
Upon measurement, a probability output is generated, where $||\alpha||^2$ as $|0\rangle$ and with a probability of $||\beta||^2$ as $|1\rangle$. The evolution of quantum bits can be described using unitary transformations acting on single or multiple qubits, where the former allows for arbitrary superposition, and the latter entangles multiple qubits into more complex quantum systems. These unique quantum characteristics grant access to an infinite-dimensional Hilbert space, which is the fundamental source of the exceptional computational power of quantum computing.

Applying a series of quantum gates to quantum bits yields a quantum circuit, and the design of any quantum computing task involves crafting a specific functional quantum circuit. NISQ quantum computers exhibit heightened sensitivity to larger circuit depths, thus emphasizing lower depths, minimal quantum gate counts, and more efficient utilization of quantum entanglement resources in quantum circuit design \cite{EisertJ}.

\subsection{Quantum Machine Learning and Parametrized Quantum Circuit}
\label{Quantum Machine Learning and Parametrized Quantum Circuit}

QML is a method of performing machine learning tasks using quantum computers. In QML, parametrized quantum circuit (PQC) is a commonly used quantum circuit structure that can adapt to various types of scenes \cite{BenedettiM}.

Similar to selecting appropriate neural network architectures in classical deep learning tasks, choosing the right PQC template is a challenging task. These templates lack any unique handling of datasets or use cases, potentially leading to circuit redundancies and impacts on parallel quantum circuit execution, which are particularly pronounced in multi-qubit tasks. Typically, the parameterized quantum gates are connected with multiple CNOT gates in a ladder-like fashion to form a parameterized layer, and multiple layers of PQCs are cascaded to enhance the expressivity of the overall ansatz.

Another prominent feature of PQC is that the learning process is reflected in the continuous adjustment of parameters, implying that quantum circuits need to be repeatedly executed with the same circuit depth a large number of times\cite{PérezSalinasA}. While some effective strategies can mitigate this issue \cite{CerezoM}, the accumulated total depth remains a key burden in large-scale learning tasks.

\subsection{Heuristic Quantum Circuit Design}
\label{Heuristic Quantum Circuit Design}

While templated PQCs offer versatility, they overlook the characteristics of datasets in task scenarios and exhibit redundancy in circuit resources. Designing quantum circuits with higher performance is one of the core tasks in quantum computing.

The formulation of quantum circuit structures embodies an optimization problem, governed by constraints such as the count of quantum physical qubits, adherence to the principles of quantum mechanics for logical consistency, evolution depth, and the composition and quantities of quantum gates. In practical contexts, careful attention must also be paid to the characteristics of the quantum dataset under examination. From this perspective, the universal variational quantum circuit templates, which disregard the content of the data, essentially undertake only an initial optimization endeavor by simplifying and overlooking numerous constraints.

Evolutionary algorithms have emerged as powerful tools for automatically generating optimal quantum circuits tailored to specific datasets. The study of evolutionary strategies has matured significantly \cite{ZhangA, KrylovG, ArufeLQuantumcircuit} finding widespread applications in diverse fields such as combinatorial optimization \cite{ArufeLNewcoding}  and image quantum processing \cite{AltaresLópezS}.

More specifically, evolutionary algorithms need to select a set of candidate quantum gates and appropriately describe the topology of the quantum circuit to adaptively design circuits. Both of these significantly influence the algorithm's performance \cite{WuW}. By strategically evolving, new individuals with higher fitness are obtained until the optimal quantum circuit is found or a predetermined endpoint is reached.

\subsection{Unsupervised Quantum Approach for Image Similarity}
\label{Unsupervised Quantum Approach for Image Similarity}

Silver et al.'s groundbreaking work \cite{SilverDSLIQ} introduced the first quantum learning approach for effective unsupervised image similarity detection on NISQ quantum computers. SliQ utilizes triplets to encode image pairs, followed by similarity learning using variational quantum circuits, and enhances model stability through a projection variance loss function, achieving promising performance on the landscape dataset. SliQ adopted an early contrastive learning strategy, achieving a cost-effective transfer to the quantum domain.

However, the process of constructing triplets in SliQ relies on predefining the Euclidean distance to specify image similarity before training begins. On one hand, performing a large number of classical image similarity calculations severely limits the potential advantages of this quantum model. On the other hand, while SliQ employs a parametrized quantum circuit as the model ansatz and adopts a multi-layered hierarchical architecture to enhance its expressive power, it still fails to fully exploit the performance advantages of quantum circuits.

\section{Quantum Unsupervised Similarity Learning}
\label{Quantum Unsupervised Similarity Learning}

Motivated by fully exploiting the advantages of quantum computing in unsupervised image similarity tasks, we propose the QUSL framework, which integrates the strengths of quantum computing and unsupervised learning. Building upon this motivation, QUSL achieves effective generation of quantum $A$, $P$, $N$-triplets without relying on classical algorithms and introduces a heuristic learning process to systematically explore the solution space for realizing high-performance quantum feature mappings.

\subsection{$A,P,N$-Triplet Construction and Quantum Embedding}
\label{$A,P,N$-Triplet Construction and Quantum Embedding}

$A,P,N$-triplets are a key concept in unsupervised learning, forming the basis for numerous mainstream self-supervised learning models \cite{VeitA,he2020momentum}. Specifically, each triplet comprises an anchor image, a positively labeled similar sample, and a negatively labeled dissimilar sample. During the learning process, these triplets are mapped to high-dimensional vectors in the embedding space, learning sample features by capturing the distance between vectors.

\begin{figure}[t]
	\hspace{0cm}
	\includegraphics[width=1.05\textwidth, trim=2.7cm 8.7cm 2cm 8cm,clip]{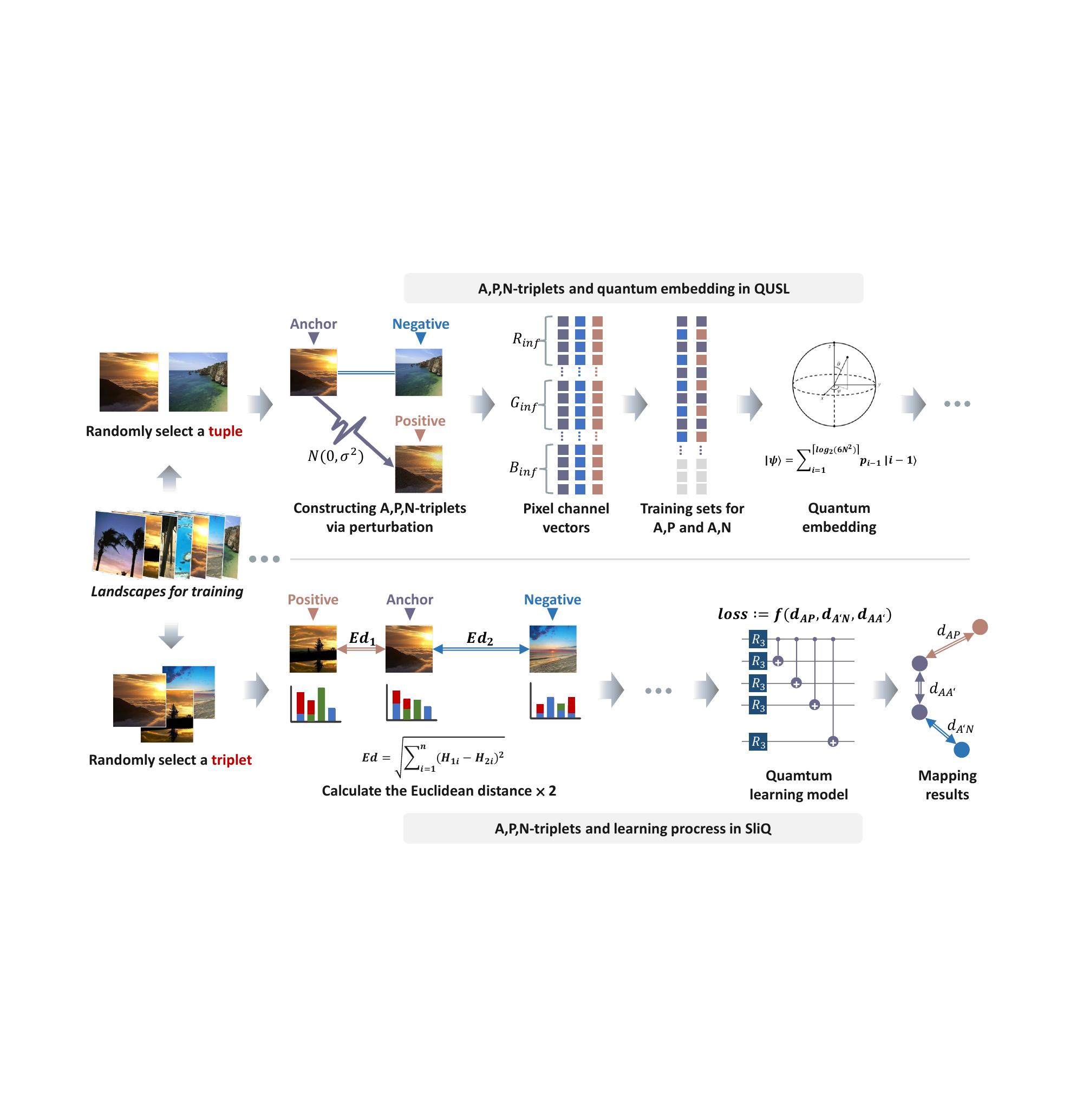}
	\caption{
		\footnotesize
		The preprocessing of $A,P,N$-triplets involves quantum embedding in QUSL. For comparison, the triplet construction and learning process of SliQ are illustrated below the figure. The specific meanings of the equations in the figure can be found in Eq.\ref{slight perturbation}, Eq.\ref{amplitude embedding}, and Eq.\ref{Ed}.}
	\label{fig2}
\end{figure}

SliQ employs triplets as a modeling process for classical image quantization, as illustrated in the lower part of Fig.\ref{fig2}. In this process, SliQ uses Euclidean distance as an evaluation metric to select positive and negative samples. This approach incurs significant additional classical costs in quantum tasks involving large-scale, high-resolution unlabeled image datasets, thus limiting the method's advantages.

QUSL uses the method of feature interweaving generated by perturbing positive images to improve mapping efficiency and eliminate the use of the oracle of visual concept. After anchor image selection, data augmentation is performed by applying slight perturbation to the anchor image as follow

\begin{equation}
	\label{slight perturbation}
	\overline{I}[c,i,j] = I[c,i,j] + \mathcal{N}(0, \sigma^2)[c,i,j],
\end{equation}

where $I[c,i,j]$ denotes the anchor image with pixel index $(i,j)$ and RGB channel information $c$, $\overline{I}$ represents the positive sample image, and $\mathcal{N}(0, \sigma^2)$ signifies Gaussian noise with a mean of $0$ and variance of $\sigma^2$. Gaussian noise, as a widely representative form of noise, exhibits stable interference effects on image features \cite{Hendrycks}, making it a universal source of perturbation. This method introduces similarity differences to the anchor image through noise perturbation, using these as positive samples. This approach draws inspiration from SimCLR's data augmentation strategy \cite{chen2020simple}, greatly reducing the cost of constructing triplets. In addition, QUSL adopts the strategy of utilizing feature interweaving to reduce the mapping process.

In the encoding phase, QUSL uses quantum amplitude embedding \cite{Rebentrost} to encode input features into quantum bits as follows

\begin{equation}
	\label{amplitude embedding}
	|\psi\rangle=\sum^{\lceil log_2 6N^2 \rceil}_{i=1}p_{i-1}|i-1\rangle,
\end{equation}
where, $\left|\psi\right\rangle$ represents the quantum state obtained by amplitude embedding a pair of training set. For an image of size $N*N$,  $p_i$ represents the $i$-th element of the training set vector, which is a normalized vector of length $6N^2$ after feature interweaving of the triplet. Thus, amplitude embedding saves a large number of quantum bits and preserves the complete image features mapped to the Hilbert space without the need for compression methods such as PCA. From the perspective of quantum information, feature interweaving of images in the triplet also results in complex entanglement between different image features, thereby augmenting the representation of image features \cite{PaineAE}. The above process is shown in the Fig.\ref{fig2}.

In summary, QUSL achieves efficient formal processing and quantumization of classical image features.

\subsection{Dataset-Oriented Quantum Learning Circuit Construction}
\label{Dataset-Oriented Quantum Learning Circuit Construction}

Heuristic quantum circuit design methodologies, such as evolutionary algorithms, can holistically address multiple constraints to determine the nearly optimal quantum state evolution process. The formulation of fitness functions for these algorithms should consider the attributes of enhanced $A, P, N$-triplets.

Unlike conventional triplets \cite{VeitA}, the quantum version's mapping procedure cannot be simply represented as the projection of three vectors onto a two-dimensional plane. This complexity stems from the incorporation of training sets, which introduces entanglement among image features, disturbing the direct correspondence between the mapped vectors and the features.

To reconcile this disparity, a corrective term must be included in the loss function for classical mapping to ensure projection coherence \cite{SilverDSLIQ}. Based on the conceptualized quantum circuit model, the training sets of $A, P, N$-triplets are embedded into the Hilbert space and encapsulated by the probability distributions obtained from measuring the first four qubits of the quantum circuit across two mappings. By combining the practical significance of fitness and the rectified mapping relationship, the fitness function, designed to maintain projection consistency, is formulated as follows
\begin{equation}
	F_{obj}:=\frac{1}{\alpha(l_{QM})+\beta\Delta},
\end{equation}
where, $\alpha$ and $\beta$ represent harmonic hyperparameters employed to determine the equilibrium of the correction term. The quantum mapping is intended to yield analogous outcomes to the classical mapping process. Therefore, $l_{QM}$ is defined as the discrepancy in vector distance as specified in the classical $A, P, N$-triplets method, formulated as
\begin{equation}
	l_{QM}=(|A_{px}-P_x|+|A_{py}-P_y|)-(|N_x-A_{nx}|+|N_y-A_{ny}|),
\end{equation}
where, $A_{px}$, $A_{py}$, $P_x$, $P_y$ denote the measure expectation of the first to fourth qubits of the training sets comprised of anchor images and positive images subsequent to their traversal through the quantum circuit. Similarly, $N_x$, $N_y$, $A_{nx}$, $A_{ny}$ signify the measure expectation of the first to fourth qubits of the training sets formed by anchor images and negative images after undergoing the quantum circuit. 

The correction term $\Delta$ is defined as the discrepancy in distance between the vectors derived from the two mappings of anchor images within the two training sets, expressed as follows
\begin{equation}
	\Delta=|A_{px}-A_{nx}|+|A_{py}-A_{ny}|.
\end{equation}
The fitness function takes into account the attributes of evolutionary algorithms and mitigates the adverse impacts resulting from the enhanced efficiency of $A, P, N$-triplets.

Designing evolutionary algorithms for quantum circuits poses challenges due to the unique characteristics of quantum circuits, such as performance symmetry and constraints in heterogeneous quantum circuits.

Performance symmetry arises from the principle of qubit permutation symmetry, where the numbering and arrangement of qubits do not affect the system's evolution operator, maintaining system equivalence. This can lead to heterogeneous quantum circuits achieving similar or identical performance, hindering population diversity in evolutionary algorithms \cite{NamY}. To mitigate this issue and maintain comprehensive coverage of the quantum circuit structure solution space, tournament selection can be employed, effectively reducing the survival rate of similar redundant individuals and balancing population diversity and convergence.

\begin{figure}[t]
	\hspace{0cm}
	\includegraphics[width=1.1\textwidth, trim=2.5cm 8.2cm 2cm 8.2cm,clip]{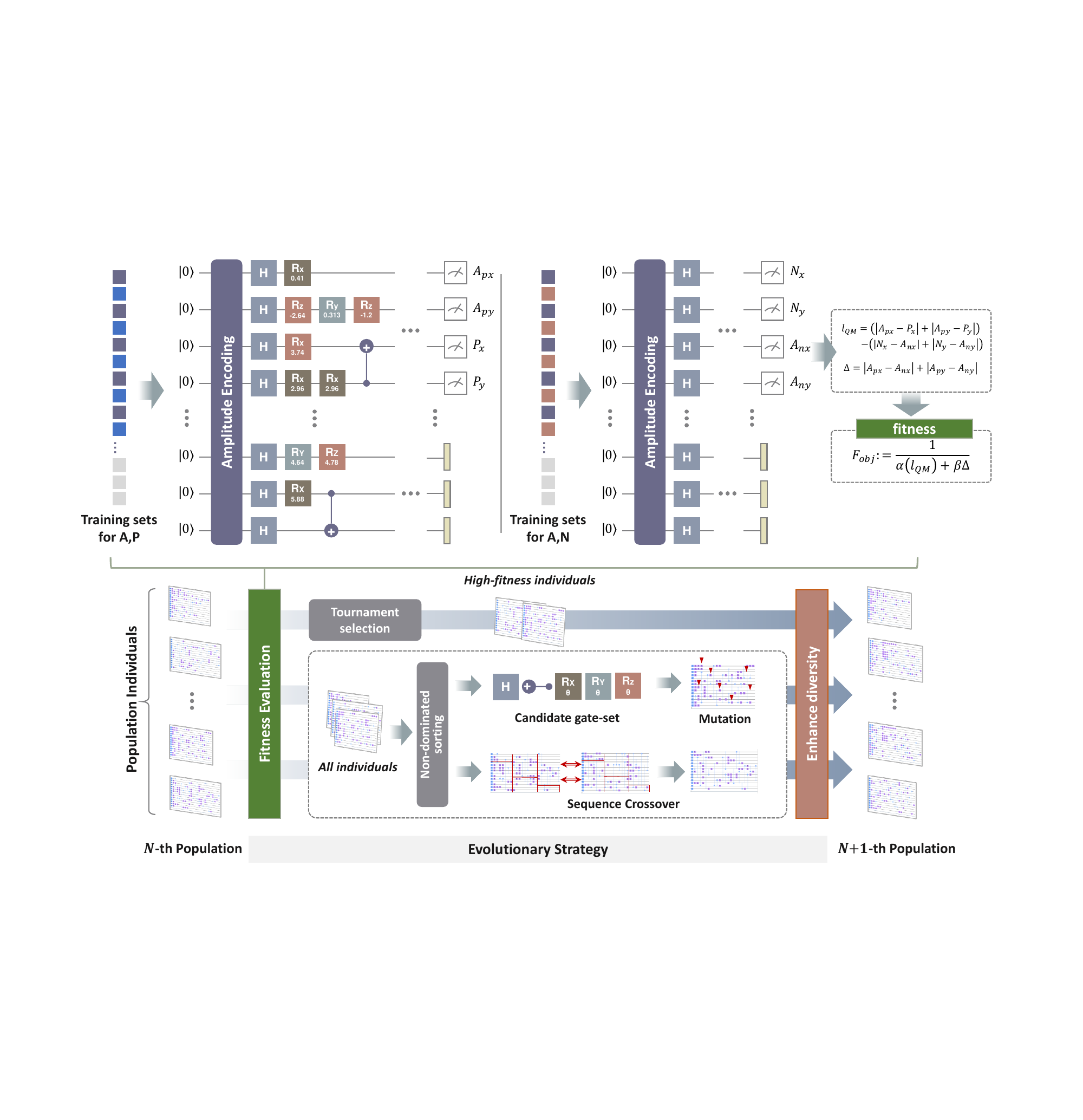}
	\caption{
		\footnotesize
		The evolutionary strategy for quantum circuits design in QUSL. To compactly illustrate the strategy, the figure does not depict the details of the components in the evolutionary algorithm.}
	\label{fig3}
\end{figure}

Performance constraints, another fundamental issue in quantum circuit design, require a holistic consideration of various aspects of quantum circuit performance evaluation. Non-dominated sorting \cite{fang2008efficient} categorizes the population based on their support, with individuals within the same category being non-dominated. This approach generates diverse trade-off solutions, focusing on the Pareto front. When applied to quantum circuit design, non-dominated sorting simplifies population complexity and prevents the generation of high-depth, complex entangled quantum circuits. The synergistic application of tournament selection and non-dominated sorting ensures expansive exploration and expedites convergence to local depth. The comprehensive evolutionary strategy is illustrated in Fig.~\ref{fig3}.

In conclusion, QUSL utilizes a comprehensive evolutionary strategy to explore high-performance quantum circuits that conform to the constraints of image similarity tasks while ensuring projection consistency in fitness settings.

\subsection{Image Similarity Performance Evaluation for Quantum Model}
\label{Image Similarity Performance Evaluation for Quantum Model}

In the domain of unsupervised image similarity assessment and similar tasks, correlation analysis is frequently employed as a primary evaluation metric \cite{MeiY}. Correlation analysis evaluates model performance by assessing the consistency between the similarity scores generated by the model and the actual similarity between images, thereby addressing the limitations associated with directly comparing score disparities among different models. This classic comparative approach exhibits high versatility and can be directly transferred to the quantum domain.

In QUSL, model performance is evaluated by computing the Spearman correlation coefficient between the similarity assessment scores outputted by the quantum model and the Euclidean distance as follow
\begin{equation}
	\label{Ed}
	Ed=\sum_{i=1}^N\sqrt{(R_{(2i)}-R_{(1i)})^2-(G_{(2i)}-G_{(1i)})^2-(B_{(2i)}-B_{(1i)})^2},
\end{equation}
where $(R_{(2i)},G_{(2i)},B_{(2i)})$ and $(R_{(1i)},G_{(1i)},B_{(1i)})$ represent the RGB values of the $i$-th pixel in the reference image and the image to be compared, respectively. In practical computations, this process is often simplified by utilizing methods such as color frequency histograms to condense pixel information for rapid calculations. In this case, the Euclidean distance can be expressed as $E_d=\sqrt{\sum_{i=1}^n(H_{1i}-H_{2i})^2}$, Here, $H_{1i}$ and $H_{2i}$ represent the channel information in the $i$-th interval of the two histograms, with $n$ denoting the total number of intervals in each histogram.

The computation of similarity assessment scores follows the following approach: The improved $A, P, N$-triplets method, which employs feature entanglement, is simplified to consider only the scenario involving anchor images and comparison images for similarity assessment. By exchanging the tuple labels of the anchor image and the positive image, this approach enables the quantification of the similarity between two images. Therefore, the similarity assessment scores $S_{sim}$ are defined as the difference between the vectors obtained from two mappings of the anchor image and the positive image after feature interweaving as follow
\begin{equation}
	S_{sim}:=\mathcal{F}(A^{(1)},A^{(2)},P^{(1)},P^{(2)}),
\end{equation}
where $A^{(i)}$ and $P^{(i)}$ represent the vectors obtained from the  anchor image and positive image, respectively, in the $i$-th mapping.  $\mathcal{F}$ denotes the operator acting on the four sets of vector  coordinates obtained during the two mapping processes, representing the  differences in coordinates between the two mappings. For simplicity,  cumulative coordinate differences can be used for computation. When the  coordinates from the two mappings are identical, $S_{sim}$ takes the  minimum value of $0$, indicating complete similarity between the anchor  image and the positive image. Conversely, when significant coordinate changes occur, $S_{sim}$ takes large positive values, with the numerical magnitude positively correlated with the degree of difference between  the anchor image and the positive image. The range of $S_{sim}$ and its  correlation with image similarity are analogous to the Euclidean  distance $Ed$, thus making it suitable for Spearman correlation  computation. Let $\mathbf{S}*{rg}$ denote the sorted collection of $S*{sim}$ series outputted by the quantum model for a sequence of $n$ pairs of  similarity detection images, and let $\mathbf{ED}_{rg}$ denote the  sorted collection of Euclidean distances $Ed$ obtained for the  corresponding image pairs sequence. Then, the Spearman correlation  coefficient $\rho$ is defined as
\begin{equation}
	\rho
	=\frac{cov(\mathbf{S}_{rg},\mathbf{ED}_{rg})}{\sigma({\mathbf{S}_{rg}})\sigma({\mathbf{ED}_{rg}})}
	= \frac{\sum_{i=1}^{n}(S_{sim(i)}-\overline{\mathbf{S}_{rg}})(Ed_{(i)}-\overline{\mathbf{ED}_{rg}})}
	{\sqrt{\sum_{i=1}^{n}(S_{sim(i)}-\overline{\mathbf{S}_{rg}})^2}\sqrt{\sum_{i=1}^{n}(Ed_{(i)}-\overline{\mathbf{ED}_{rg}})^2}},
\end{equation}
where, $S_{sim(i)}$ and $Ed_{(i)}$ denote the $i$-th elements of $\mathbf{S}_{rg}$ and $\mathbf{ED}_{rg}$, respectively, while $\overline{\mathbf{S}_{rg}}$ and $\overline{\mathbf{ED}_{rg}}$ represent the mean values of elements in the two sets. The Spearman correlation coefficient, by definition, ranges from $-1$ to $1$, with $0$ indicating no correlation, $1$ denoting perfect positive correlation, and $-1$ representing perfect negative correlation. Models exhibiting higher positive correlation values demonstrate relatively better performance. In subsequent experiments and comparative analyses, this coefficient will be utilized to assess the performance of the quantum image similarity model.

\begin{figure}[t]
	\hspace{0cm}
	\includegraphics[width=1\textwidth, trim=2cm 7.9cm 2cm 7.98cm,clip]{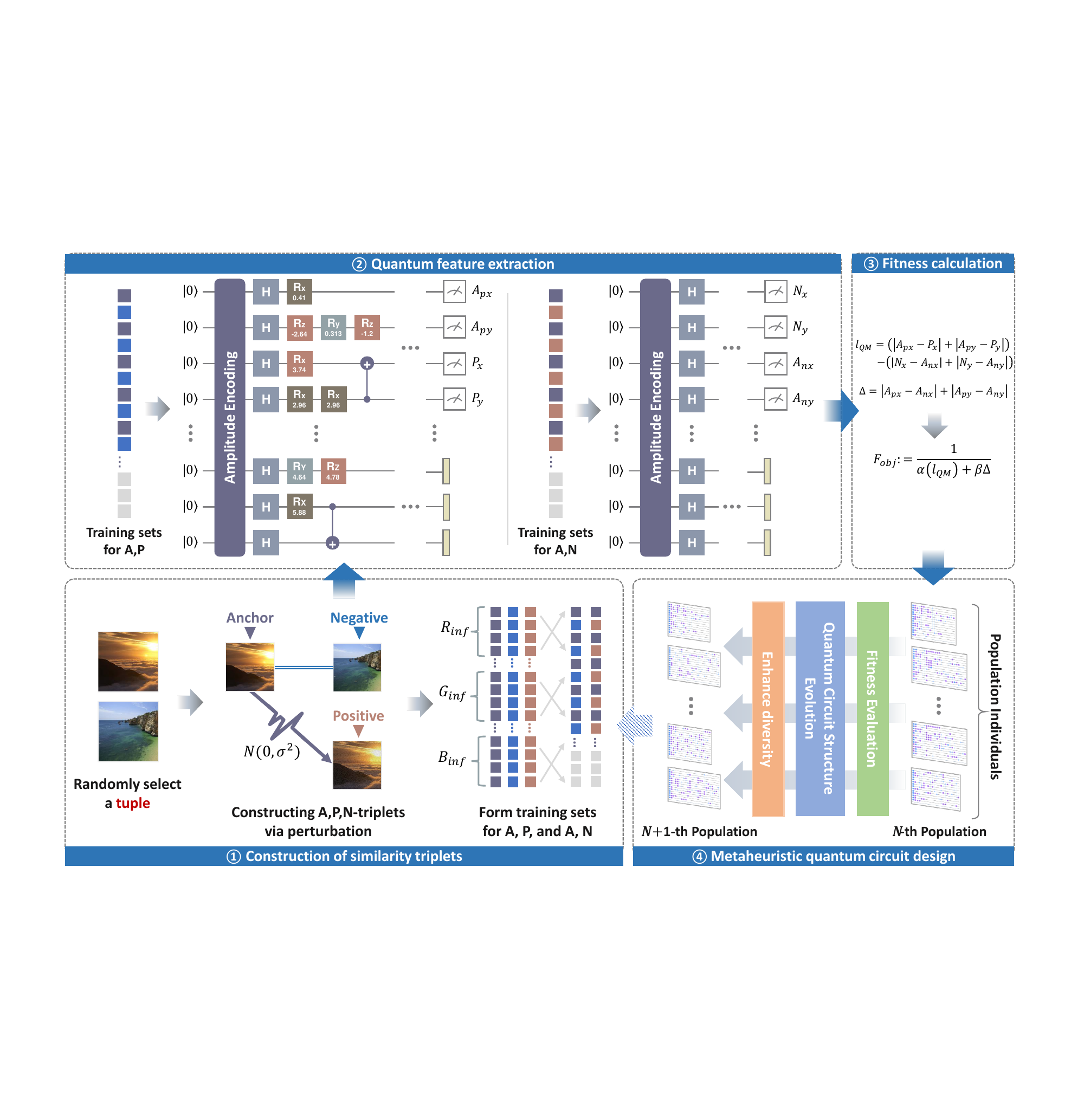}
	\caption{
		\footnotesize
		The complete process of QUSL. Some processes are summarized to ensure the clarity of the legend.}
	\label{fig_C}
\end{figure}

The complete process of QUSL is shown in Fig.~\ref{fig_C}.

\section{Experiments and Result}
\label{Experiments and Result}
This section presents an experimental evaluation of the QUSL model, demonstrating its effectiveness, robustness, and superiority compared to state-of-the-art quantum methods.

\subsection{Experiment Framework}
\label{Experiment Setting and Framework}

In the domain of unsupervised image similarity tasks, we have selected five datasets for training and testing purposes. Table \ref{Datasets Table} outlines the essential details of these datasets and the experimental configurations.

\begin{table}[htb]
	\renewcommand{\arraystretch}{1.2}
	\caption{Datasets with basic setup used.}
	\vspace{-\abovecaptionskip}
	\vspace{0.1cm}
	\label{Datasets Table}
	\centering
	\begin{threeparttable}
		\scriptsize
		\begin{tabularx}{\textwidth}{>{\centering\arraybackslash}X >{\centering\arraybackslash}X >{\centering\arraybackslash}X >{\centering\arraybackslash}X}
			\toprule
			Dataset & Categories Used & Patch Size & Qubit Used \\[-0.1cm]
			\midrule
			landscape & $7$ object categories\textsuperscript{a} & $80\times80\times3$ & $14$ \\
			COCO & $80$ object categories & $ 50\times50\times3$ & $14$ \\
			DISC21 & $21$ object categories & $ 50\times50\times3$ & $14$ \\
			CIFAR\_10 & $10$ object categories & $32\times32\times3$ & $14$ \\
			ImageNet  & $1000$ object categories & $50\times50\times3$ & $14$ \\
			\bottomrule
		\end{tabularx}
		\begin{tablenotes}
			\footnotesize
			\item[a] All datasets do not utilize existing labels in the experiments.
		\end{tablenotes}
	\end{threeparttable}
\end{table}

The Flickr landscape \cite{MenY}, comprising unlabeled colorful images depicting diverse natural landscapes, along with the label-removed COCO \cite{LinTY} and CIFAR$\_$10 \cite{krizhevsky2009learning}, serve to initially examine QUSL's adaptability to unsupervised tasks. Experiments on ImageNet \cite{deng2009imagenet} further test QUSL's performance. DISC21 \cite{DouzeM}, specifically designed for image similarity detection tasks, is instrumental in evaluating QUSL's applicability in real-world scenarios, particularly in social media contexts.

The experimental setup was configured using Python3, incorporating a hybrid combination of the Qiskit and MindSpore \cite{mq_2021} frameworks. Numerical simulations were executed on a workstation (CPU: Intel I9 9900k, GPU: GTX3090). Guided by pre-experiment instructions, the evolutionary algorithm featured a population size of $20$ and a maximum evolution generation of $20$. The gate set constituting the quantum circuit comprised ${R_X, R_Y, R_Z, H, CNOT}$. Images within the training set underwent uniform processing to ensure consistent dimensions. Following the selection of anchor images, random images from the training set were chosen to form $A, P, N$-triplets. For all models, $1000$ similarity tests were conducted, and model performance was assessed using correlation detection.

\subsection{Evaluation and Analysis}
\label{Evaluation and Analysis}

To determine the optimal perturbation level for the QUSL algorithm, we conducted a series of preliminary experiments on the landscape dataset, evaluating the algorithm's performance across different perturbation levels, characterized by the parameter $\sigma$.

\begin{figure}[t]
	\centering
	\includegraphics[width=0.75\textwidth, trim=0cm 0.1cm 0cm 0cm,clip]{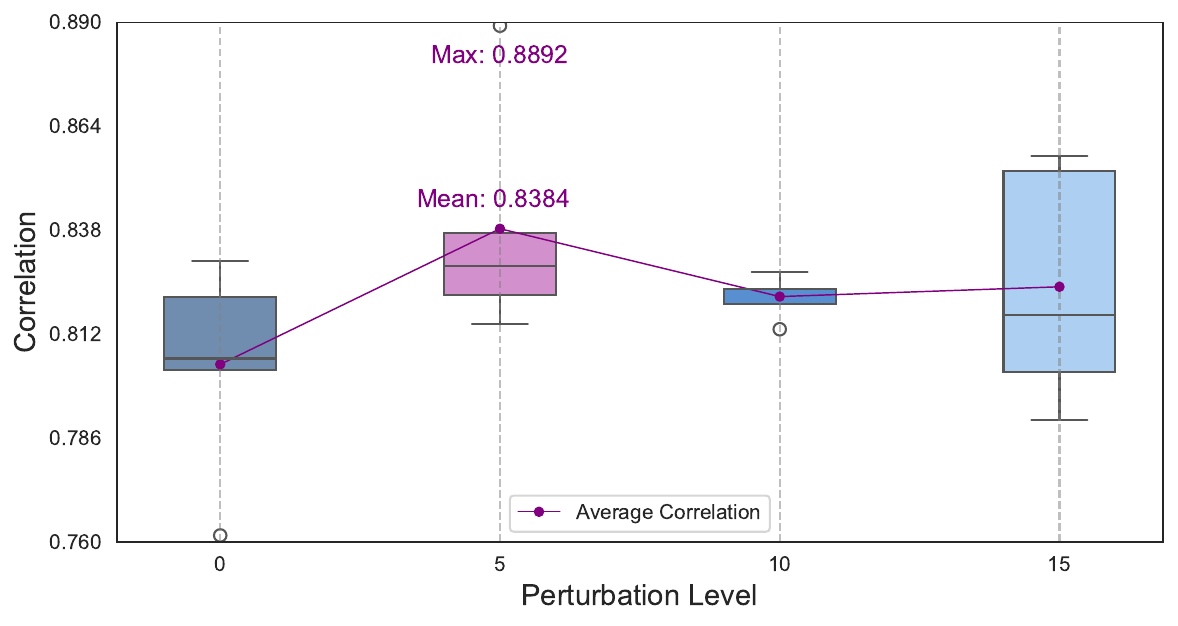}
	\caption{
		\footnotesize
		Performance of the QUSL on the landscape dataset under varying levels of perturbation.}
	\label{boxplot}
\end{figure}

Five independent experimental runs were carried out at each perturbation level, with the results presented in Fig.\ref{boxplot} and Table.\ref{varying levels perturbation}.

\begin{table}[h]
	\caption{Performance of the QUSL under varying levels of perturbation.}
	\vspace{-\abovecaptionskip}
	\vspace{0.1cm}
	\label{varying levels perturbation}
	\renewcommand{\arraystretch}{1.2}
	\centering
	\begin{threeparttable}
		\scriptsize
		\begin{tabularx}{\textwidth}{*{6}{>{\centering\arraybackslash}X} >{\centering\arraybackslash}p{2.5cm}}
			\toprule
			Level & \multicolumn{5}{c}{Correlation performance$^a$} & Mean value \\[-0.1cm]
			\midrule
			$0$ & $\textbf{0.830}^b$ &$0.762$ &$0.821$ &$0.803$ &$0.806$ &$0.804(\pm 0.024)^c$\\
			$5$ & $0.837$ &$0.829$ &$\textbf{0.889}$ &$0.814$ &$0.822$ & \textbf{0.838}($\pm \textbf{0.026}$) \\
			$10$ & $0.823$ &$\textbf{0.828}$ &$0.820$ &$0.819$ &$0.823$ &$0.823(\pm 0.024)$  \\
			$15$ & $0.817$ &$0.791$ &$0.803$ &$0.853$ &$\textbf{0.857}$ &$0.824(\pm 0.025)$  \\
			\bottomrule
		\end{tabularx}
		\begin{tablenotes}
			\footnotesize
			\item[a] Five independent training and testing sessions were conducted at each level.
			\item[b] The bold numbers indicate the highest correlation achieved in each experimental group.
			\item[c] "$\pm$" represents the standard deviation, reflecting the dispersion of sample data, as it does elsewhere in this paper. 
		\end{tablenotes}
	\end{threeparttable}
\end{table}

In comparative experiments conducted under identical experimental conditions, when the perturbation level was set to $5$ (indicated by purple shading in the Fig.~\ref{boxplot}), the QUSL model achieved the best training performance of $0.8892$, exhibiting the highest average training effectiveness and acceptable data dispersion. In contrast, the performance under other conditions showed lower training results and instability. Furthermore, the perturbation level has a significant impact on model performance, resulting in a performance discrepancy of up to $12\%$ in terms of correlation.

Therefore, by comprehensively considering relevant performance metrics, including maximum performance, average performance, and robustness, a perturbation level of $5$ can be selected as the parameter setting for subsequent experiments. Moreover, the perturbation level can serve as a controllable threshold for image similarity during the learning process, or an adaptive approach can be employed to more precisely regulate the perturbation level.

\begin{figure}[t]
	\centering
	\includegraphics[width=1\textwidth, trim=0cm 0.2cm 0cm 0.25cm, clip]{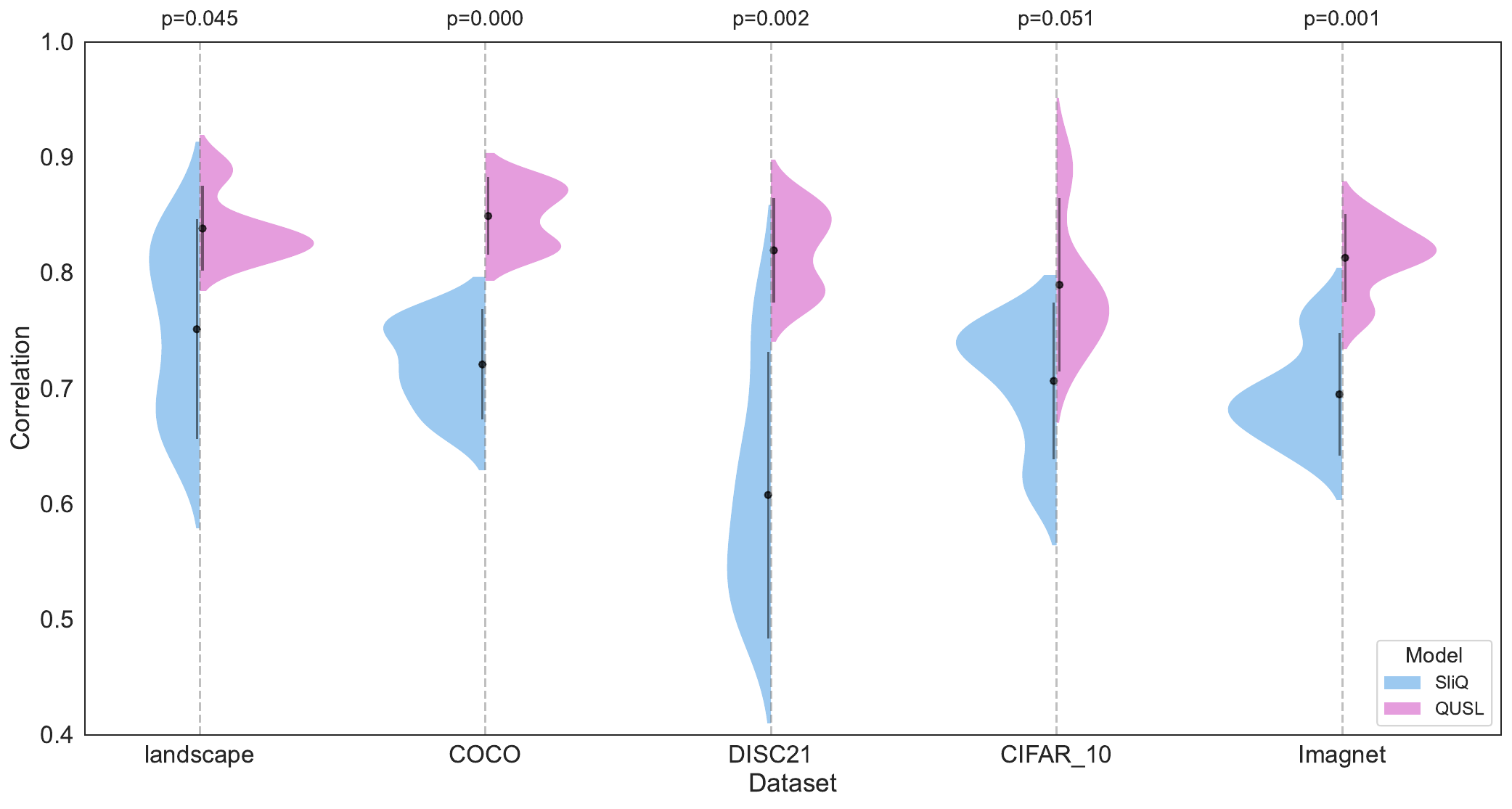}
	\caption{
		\footnotesize
		Performance evaluation of QUSL and SliQ across different datasets. In the violin plot, the blue and purple areas represent the distribution states of correlation obtained by SliQ and QUSL, respectively. Error bars represent $95\%$ confidence intervals}
		\label{violin}
\end{figure}

Controlled experiment were designed to comprehensively examine the performance efficacy and robustness of QUSL on five datasets, while concurrently benchmarking against SliQ. QUSL and SliQ underwent five training sessions and correlation tests of equal scale on each of the five datasets, recording the best results obtained during the training process, which are presented in Fig.~\ref{violin} and Table.~\ref{tab:performance_comparison}.

\begin{table}[t]
	\caption{Performance comparison between SliQ and QUSL based on five independent experiments.}
	\label{tab:performance_comparison}
	\centering
	\scriptsize
	\begin{threeparttable}
		\renewcommand{\arraystretch}{1.2}
		\setlength{\tabcolsep}{4pt}
		\begin{tabular*}{\textwidth}{@{\extracolsep{\fill}}lcccccccc@{}}
			\toprule
			Dataset & Model & 1st & 2nd & 3rd & 4th & 5th & Mean value & Improv$^a$. \\
			\midrule
			\multirow{2}{*}{landscape} 
			& SliQ & 0.835 & 0.814 & 0.758 & 0.692 & 0.657 & 0.751 ($\pm$0.078) & \multirow{2}{*}{$\uparrow$8.7\%} \\
			& QUSL & \textbf{0.889} & 0.837 & 0.829 & 0.822 & 0.814 & \textbf{0.838} ($\pm$\textbf{0.031}) & \\
			\addlinespace[0.8em]
			\multirow{2}{*}{COCO}
			& SliQ & 0.757 & 0.757 & 0.722 & 0.699 & 0.668 & 0.721 ($\pm$0.035) & \multirow{2}{*}{$\uparrow$12.8\%} \\
			& QUSL & \textbf{0.876} & 0.875 & 0.852 & 0.822 & 0.821 & \textbf{0.849} ($\pm$\textbf{0.027}) & \\
			\addlinespace[0.8em]
			\multirow{2}{*}{DISC21}
			& SliQ & 0.758 & 0.649 & 0.593 & 0.592 & 0.528 & 0.624 ($\pm$0.096) & \multirow{2}{*}{$\uparrow$19.5\%} \\
			& QUSL & \textbf{0.861} & 0.847 & 0.825 & 0.787 & 0.777 & \textbf{0.819} ($\pm$\textbf{0.037}) & \\
			\addlinespace[0.8em]
			\multirow{2}{*}{CIFAR\_10}
			& SliQ & 0.743 & 0.742 & 0.742 & 0.686 & 0.619 & 0.706 ($\pm$0.052) & \multirow{2}{*}{$\uparrow$11.9\%} \\
			& QUSL & \textbf{0.890} & 0.792 & 0.769 & 0.766 & 0.731 & \textbf{0.790} ($\pm$\textbf{0.059}) & \\
			\addlinespace[0.8em]
			\multirow{2}{*}{Imagnet}
			& SliQ & 0.761 & 0.706 & 0.681 & 0.647 & 0.679 & 0.695 ($\pm$0.042) & \multirow{2}{*}{$\uparrow$17.0\%} \\
			& QUSL & \textbf{0.848} & 0.824 & 0.818 & 0.809 & 0.765 & \textbf{0.813} ($\pm$\textbf{0.030}) & \\
			\bottomrule
		\end{tabular*}
		\begin{tablenotes}
			\footnotesize
			\item[a] Improvement of QUSL(ours) over SliQ in terms of mean correlation performance.
		\end{tablenotes}
	\end{threeparttable}
\end{table}

In all five datasets, QUSL demonstrated significant enhancements in both performance and stability. In comparison to SliQ, QUSL exhibited a remarkable improvement in correlation, with increases ranging from a minimum of $8.7\%$ to a maximum of $19.5\%$. Regarding stability, QUSL displayed a distinct advantage over SliQ in terms of the standard deviation of performance means, particularly evident in the DISC21 dataset, where SliQ's performance exhibited substantial fluctuations, while QUSL's stability remained largely unaffected by the application scenarios.

QUSL's performance improvements and high robustness can be primarily attributed to its ability to adapt to dataset characteristics through evolutionary algorithms. Conversely, the parameter adjustment of template-based variational quantum circuits may encounter difficulties in capturing the intrinsic properties of the dataset, resulting in performance degradation under equivalent training conditions. This set of experiments directly demonstrates QUSL's performance in scenarios close to real-world applications and showcases its robustness across different types of image tasks.

A series of experiments were designed to further investigate the detailed training process of QUSL, showcasing its accuracy and interpretability.

\begin{figure}[t]
	\centering
	\includegraphics[width=1\textwidth, trim=2.3cm 0.5cm 0cm 1.8cm,clip]{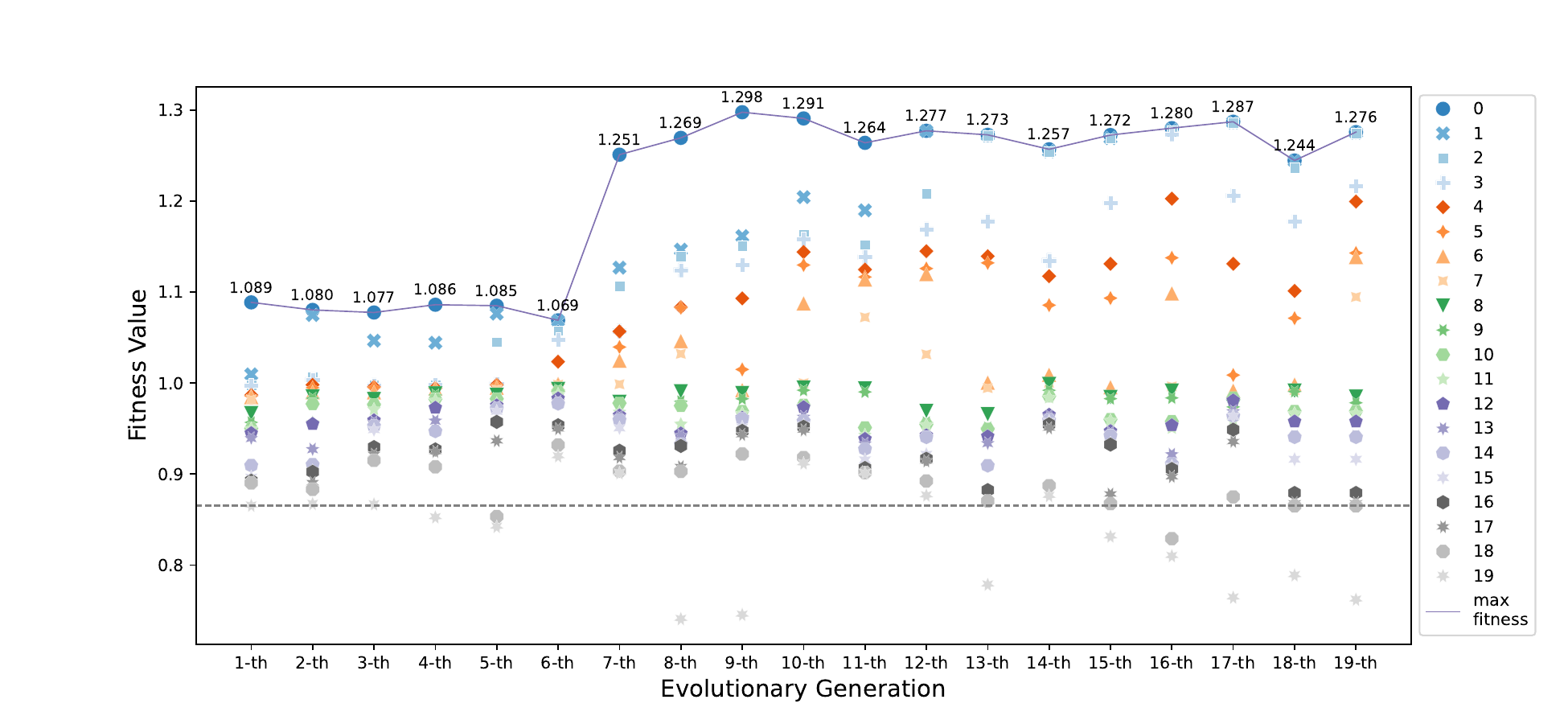}
	\caption{
		\footnotesize
		Fitness trajectory of individuals across generations during the evolutionary process in landscape. The figure annotates the fittest individuals in each generation of $20$ individuals with corresponding values. The dashed line below the image represents the lowest fitness among individuals in the first generation. Given the tendency of evolutionary algorithms to converge towards optimization, individuals below the dashed line are eliminated during the evolution process.}
	\label{Evolution_line_chart}
\end{figure}

Fig.\ref{Evolution_line_chart} illustrates the fluctuation of fitness values for all individuals within one evolutionary cycle. It is important to note that to ensure broad adaptability of individuals to all image combinations within the dataset, QUSL employs random sampling of the dataset during the training process, assembling multiple sets for fitness evaluation. This approach leads to the stabilization of the best individual's fitness within a certain range, rather than strictly increasing monotonically and ultimately converging during the evolution process. It is evident that as the number of generations increases, the fitness of the best individual improves, demonstrating the success of the evolutionary process. 

It is worth noting that during the evolution process, there are still individuals with significantly different quantum circuit structures but no apparent difference in fitness performance. A typical example of this scenario is demonstrated in \ref{appendixB}. This further demonstrates the threat of low population diversity in metaheuristic quantum circuit design due to the inherent characteristics of quantum circuits.

\begin{figure}[t]
	\centering
	\includegraphics[width=0.9\textwidth, trim=0cm 0.5cm 0cm 0cm,clip]{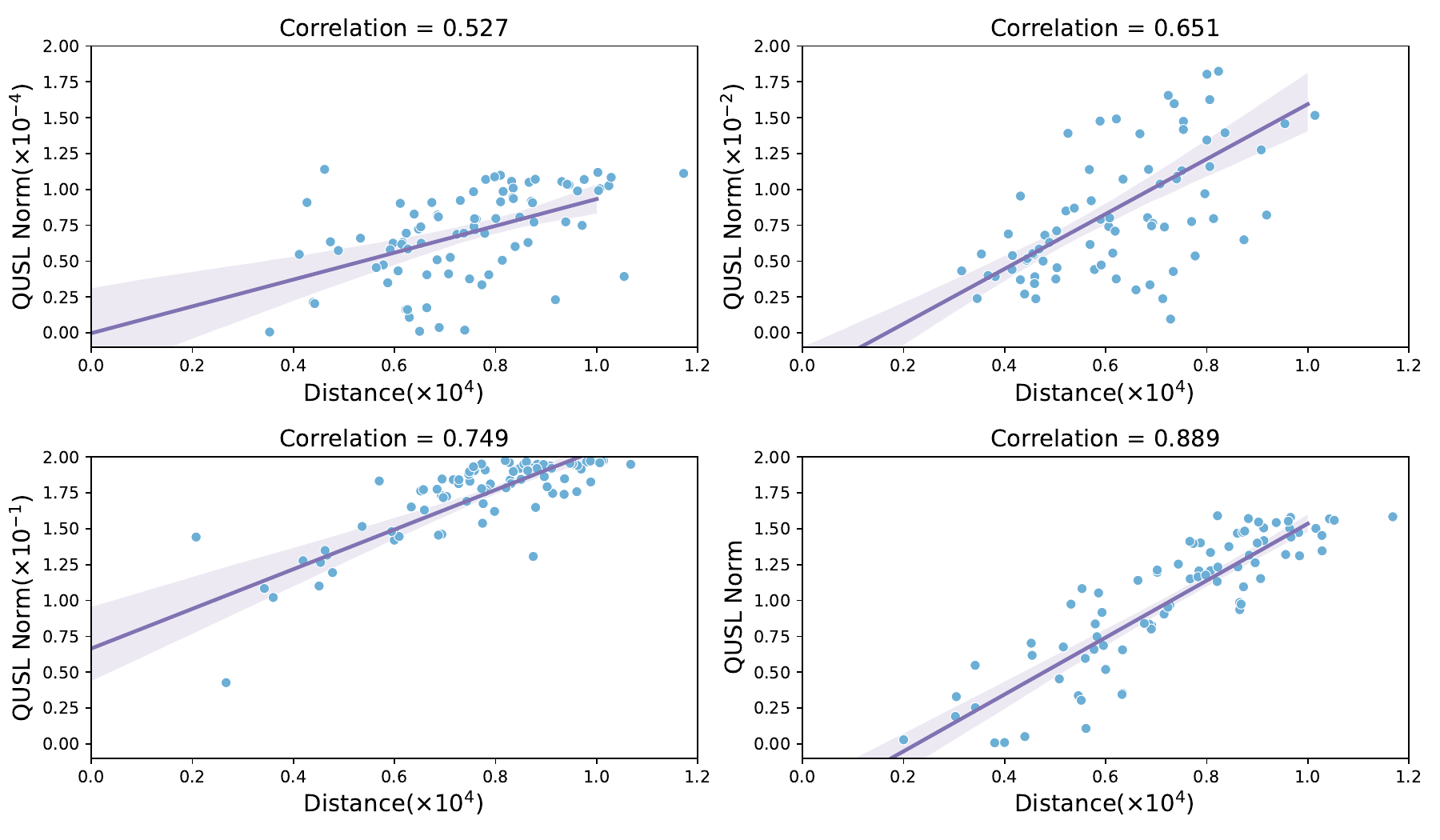}
	\caption{
		\footnotesize
		Relationship between euclidean distance and QUSL norm, the output of method, across different correlation levels. The purple line represents the regression analysis and corresponding confidence interval of the scatter plot, directly reflecting the correlation between the similarity of QUSL and the Euclidean distance.}
		\label{Correlation_of_4_times}
\end{figure}

The validity of using correlation as a model performance metric can be verified by the correspondence between the distribution of image similarities and their correlation. The results of this set of experiments are presented in Fig.\ref{Correlation_of_4_times}, with the validation conducted on the landscape dataset.

\begin{figure}[H]
	\centering
	\includegraphics[width=0.75\textwidth, trim=0cm 0.13cm 0cm 0.9cm,clip]{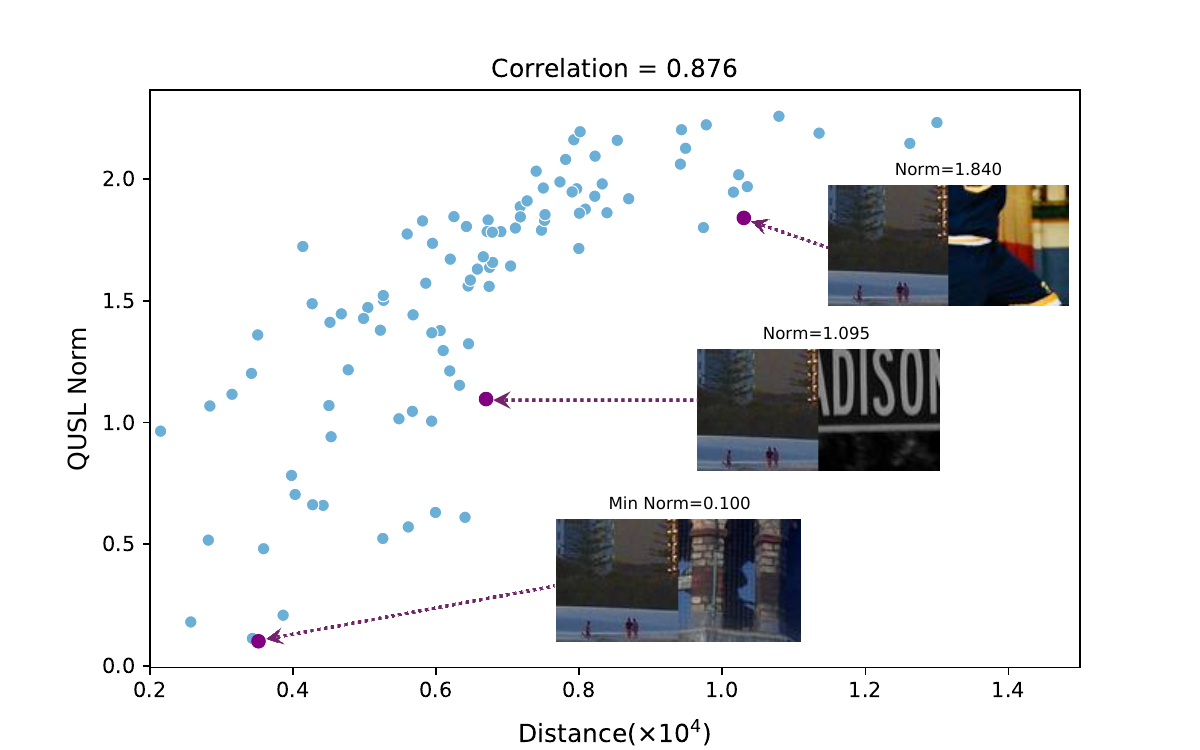}
	\caption{
		\footnotesize
		Results of QUSL on the COCO. The model's correlation is $0.876$, achieved by the optimal individual during the evolution process. When QUSL norm reaches its minimum value of $0.1$, QUSL identifies the most similar images in the current test set.}
	\label{coco}
\end{figure}

\begin{figure}[H]
	\centering
	\includegraphics[width=0.75\textwidth, trim=0cm 0.13cm 0cm 0.9cm,clip]{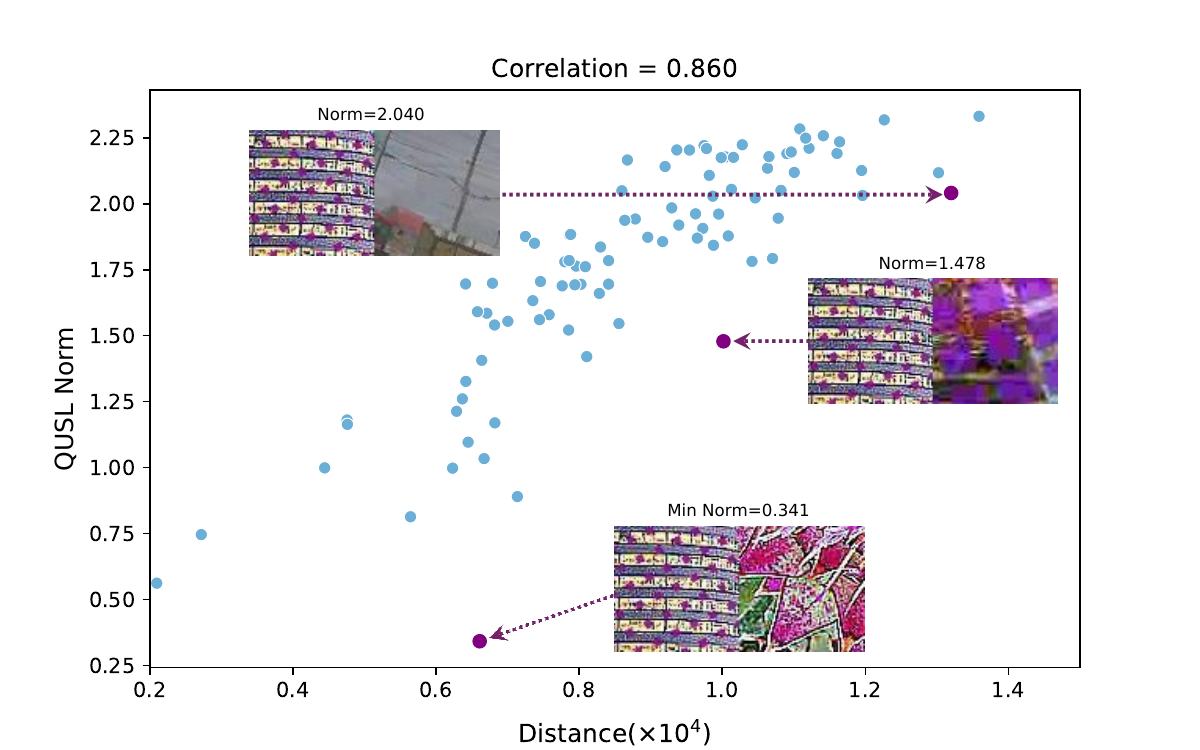}
	\caption{
		\footnotesize
		Results of QUSL on the DISC21. The model's correlation is $0.860$. When QUSL norm reaches its minimum value of $0.341$, QUSL identifies the most similar images in the current test set.}
	\label{Dis21}
\end{figure}

As the number of generations increases, the correlation between QUSL norm and Euclidean distance also gradually improves, reaching a maximum correlation of $0.889$ in this evolution, which is a significant advantage compared to SliQ. To more clearly demonstrate the correctness of the correlation metric, Fig.\ref{coco} and Fig.\ref{Dis21} respectively illustrate the true distribution of images in the models with the best correlation on the COCO and DISC21 datasets.

As illustrated in the figures, the Euclidean distance and QUSL norm exhibit a positive correlation, and images with smaller QUSL norm values demonstrate more significant visual similarity. This reflects the alignment between the image similarity determined by Euclidean distance and the image similarity assessed by QUSL.

It is crucial to highlight that the test set is randomly sampled from the dataset, and consequently, there may be an absence of image combinations that are visually indistinguishable to human perception. Nevertheless, the fluctuations in the similarity values generated by QUSL can still offer valuable insights into the variations in visual similarity among the images. The aforementioned experiments collectively validate the appropriateness of employing correlation as a model performance evaluation metric and further corroborate the model's correctness.

\begin{figure}[H]
	\centering
	\includegraphics[width=1\textwidth, trim=0cm 0.3cm 0cm 0.2cm,clip]{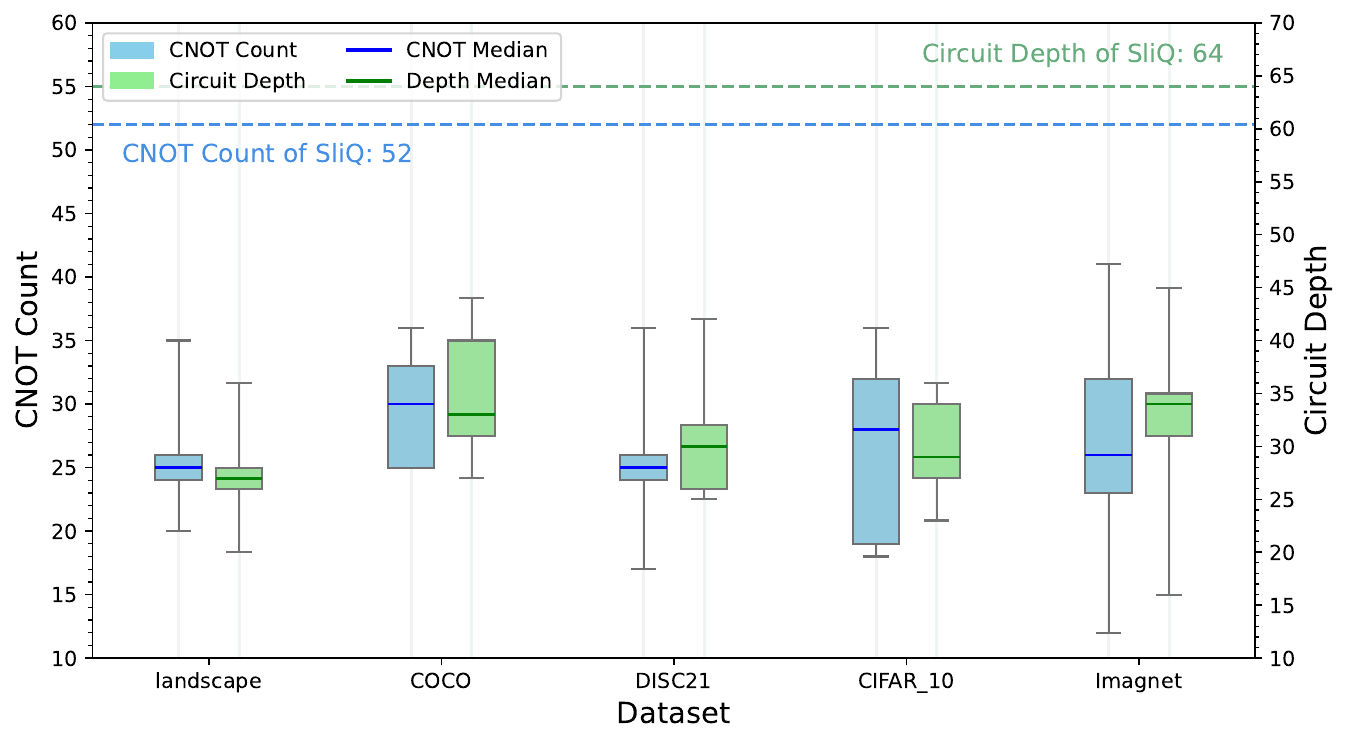}
	\caption{
		\footnotesize
		CNOT count and circuit depth utilized in the quantum circuit of QUSL. The green and blue dashed lines above the plots represent the circuit depth and CNOT count, respectively, of the variational quantum circuit templates used by SliQ, which remain unchanged across different datasets.}
	\label{CNOT_and_Deep}
\end{figure}

In terms of core quantum resources, a set of numerical experiments and experiments on quantum computers were designed to evaluate the comparison between QUSL and SliQ in terms of quantum circuit depth and CNOT count. Fig.\ref{CNOT_and_Deep} and Table.\ref{CNOT and circuit depth} illustrates a comparison of the quantum circuit performance between the two approaches.

\begin{table}[H]
	\caption{CNOT count and circuit depth utilized in the quantum circuit of QUSL.}
	\label{CNOT and circuit depth}
	\centering
	\scriptsize
	\begin{threeparttable}
		\renewcommand{\arraystretch}{1.2}
		\setlength{\tabcolsep}{4pt}
		\begin{tabular*}{\textwidth}{@{\extracolsep{\fill}}lcccccccc@{}}
			\toprule
			Dataset & Metric & 1st & 2nd & 3rd & 4th & 5th & Mean value$^a$ & Improv$^b$ \\
			\midrule
			\multirow{2}{*}{landscape} 
			& CNOT count & 35 & \textbf{20}$^c$ & 26 & 24 & 25 & 26.0 ($\pm$4.9) & $\downarrow$\textbf{50.0\%} \\
			& Circuit depth & 36 & \textbf{20} & 28 & 27 & 26 & 27.4 ($\pm$5.1) & $\downarrow$\textbf{57.2\%} \\
			\addlinespace[0.8em]
			\multirow{2}{*}{COCO}
			& CNOT count & 30 & 33 & 36 & \textbf{25} & 25 & 29.8 ($\pm$4.2) & $\downarrow$\textbf{42.7\%} \\
			& Circuit depth & 33 & 40 & 44 & \textbf{27} & 31 & 35.0 ($\pm$6.2) & $\downarrow$\textbf{45.3\%} \\
			\addlinespace[0.8em]
			\multirow{2}{*}{DISC21}
			& CNOT count & \textbf{17} & 24 & 25 & 36 & 26 & 25.6 ($\pm$6.3) & $\downarrow$\textbf{50.8\%} \\
			& Circuit depth & 26 & 30 & \textbf{25} & 42 & 32 & 31.0 ($\pm$6.1) & $\downarrow$\textbf{51.6\%} \\
			\addlinespace[0.8em]
			\multirow{2}{*}{CIFAR\_10}
			& CNOT count & 32 & 19 & 28 & \textbf{18} & 36 & 26.6 ($\pm$7.1) & $\downarrow$\textbf{48.8\%} \\
			& Circuit depth & 36 & \textbf{23} & 34 & 27 & 29 & 29.8 ($\pm$4.7) & $\downarrow$\textbf{53.4\%} \\
			\addlinespace[0.8em]
			\multirow{2}{*}{Imagnet}
			& CNOT count & 41 & \textbf{12} & 23 & 32 & 26 & 26.8 ($\pm$9.7) & $\downarrow$\textbf{48.5\%} \\
			& Circuit depth & 45 & \textbf{16} & 31 & 34 & 35 & 32.2 ($\pm$9.3) & $\downarrow$\textbf{49.7\%} \\
			\bottomrule
		\end{tabular*}
		\begin{tablenotes}
			\footnotesize
			\item[a] The CNOT count and circuit depth are both integers, mean value merely reflects the relative performance of the corresponding quantum circuits.
			\item[b] Improvement of QUSL(ours) over SliQ in terms of  the decrease ratio of the average CNOT count and circuit depth compared to their respective values in SliQ.
			\item[c] Bold text indicates the best-performing parameters within a set of experiments.
		
		\end{tablenotes}
	\end{threeparttable}
\end{table}

Across all datasets used, the quantum circuits derived through QUSL exhibit reductions of around $50\%$ in both CNOT counts and circuit depth. This indicates that within acceptable fluctuations, the quantum circuits generated by QUSL significantly outperform SliQ in key performance indicators, directly contributing to the model's advantage in quantum resource utilization. The demonstration of high-performance quantum circuits is presented in \ref{appendixA}.

Validation experiments on quantum computers were designed to examine the potential advantages of QUSL in the NISQ era. The experiments were conducted on the $ibm_osaka$ quantum computer with $500$ shots. Table \ref{realQcomputer} presents a comparison of the running times between the high-performance quantum circuits used in QUSL and the template circuits employed in SliQ.

\begin{table}[t]
	\caption{Comparison of the running times of optimal quantum circuits on real quantum computers.}
	\vspace{-\abovecaptionskip}
	\vspace{0.1cm}
	\label{realQcomputer}
	\renewcommand{\arraystretch}{1.2}
	\centering
	\scriptsize
	\begin{threeparttable}
		\begin{tabular}{ccccccc}
			\toprule
			Dataset & \multicolumn{2}{c}{landscape} & \multicolumn{2}{c}{COCO} & \multicolumn{2}{c}{DISC21}\\[-0.1cm]
			\cmidrule(lr){2-3} \cmidrule(lr){4-5} \cmidrule(lr){6-7}
		Run time$(s)$$^a$ & \makecell{SliQ} & \makecell{QUSL} & \makecell{SliQ} & \makecell{QUSL} & \makecell{SliQ} & \makecell{QUSL} \\[-0.1cm]
			\midrule
			1 & $1.64$ & $1.55$ & $1.60$ & $1.50$ & $1.67$ & $1.48$ \\
			2 & $1.73$ & $1.59$ & $1.73$ & $1.61$ & $1.70$ & $1.49$ \\
			3 & $1.76$ & $1.60$ & $1.76$ & $1.72$ & $1.70$ & $1.51$ \\
			4 & $1.21$ & $1.64$ & $1.76$ & $1.72$ & $1.74$ & $1.65$ \\
			5 & $1.90$ & $1.80$ & $1.80$ & $1.73$ & $1.74$ & $1.75$ \\
			\midrule
			\makecell{Mean value}& \makecell{$1.77^c$\\$(\pm 0.089)$} & \makecell{$1.64$\\$(\pm0.087)$} & \makecell{$1.77$\\$(\pm 0.080)$} & \makecell{$1.66$\\$(\pm0.089)$} & \makecell{$1.71$\\$(\pm0.027)$} & \makecell{$1.58$\\$(\pm0.109)$} \\ 
			\makecell{Optimization}&-&$\downarrow\textbf{0.13}$&-&$\downarrow\textbf{0.11}$&-&$\downarrow\textbf{0.13}$\\
			\bottomrule
		\end{tabular}
		\begin{tablenotes}
			\footnotesize
			\item[a] Five independent running times sorted in ascending order.
		\end{tablenotes}
	\end{threeparttable}
\end{table}

Through five independent circuit runs, it can be observed that QUSL consistently reduces the running time of individual quantum circuits. The time advantage at the single-circuit level is significantly amplified by the large number of circuit runs during the training process, which will be reflected in the model's training efficiency and practical application efficiency. This advantage of QUSL becomes more evident in large-scale tasks.

Through numerical simulations and a series of experiments on quantum computers, the QUSL algorithm demonstrates robust performance improvements in image similarity tasks, achieving higher average correlations across multiple datasets. In challenging datasets closer to real-world scenarios, such as DISC21, QUSL exhibits higher performance and stronger stability. More importantly, compared to SliQ's variational template, QUSL significantly reduces quantum resource consumption. Combined with validation on quantum computers, QUSL demonstrates higher learning efficiency in NISQ environments relative to SliQ.

Overall, QUSL, as an unsupervised quantum image similarity model, surpasses current state-of-the-art quantum image similarity detection methods in terms of accuracy, robustness, and quantum resource efficiency.

\section{Discussion and Futher Work}
\label{Discussion and Futher Work}

Along with our work, SliQ's \cite{SilverDSLIQ} performance demonstrates that quantum computing can be applied to image processing tasks with potential efficiency gains. Due to NISQ limitations, quantum machine learning algorithms native to quantum computers are challenging to implement. Both QUSL and SliQ represent quantum transfers of classical unsupervised image tasks, particularly contrastive learning tasks. In comparison, QUSL goes further than the latest methods in realizing a quantum implementation of the classical triplet method. Furthermore, this work reveals the importance of quantum circuit design in quantum unsupervised learning tasks. As an efficient method, heuristic quantum circuit architecture search can design high-performance circuits that adapt to task requirements. This directly alleviates the dependence of complex unsupervised tasks on deep, parameter-rich PQCs, making it more suitable for the NISQ era. This work provides a preliminary exploration of large-scale unsupervised contrastive learning in the quantum domain.

These methods utilize the correlation between the computed model loss and classical similarity norm to evaluate model performance. This ingenious and transferable approach addresses the challenge of objectively assessing image similarity in unlabeled datasets, proving effective within the limited application scenarios addressed in our work. However, this evaluation method remains open to refinement. For instance, exploring more suitable correspondences between losses and classical image similarity detection methods within different numerical ranges. This refinement process stands as a significant future endeavor within our research trajectory.

In the process of constructing A,P,N-triplets, noise is used for data augmentation. The efficacy of this strategy was validated through subsequent experimental analysis. While noise is conventionally perceived as an impediment in both classical and quantum contexts — noise cuts both ways — it can also serve as a valuable asset for specific tasks \cite{LiuD}. Although noise mitigation techniques have been extensively explored in quantum computing, acknowledging noise as a potentially beneficial factor paves the way for novel approaches and methodologies

\section{Conclusion}
\label{Conclusion}
The QUSL method proposed in this work focuses on quantum unsupervised image similarity detection tasks. QUSL employs heuristic methods to adapt quantum circuits to dataset characteristics and utilizes noise-based data augmentation for triplet construction. A series of numerical simulations and quantum computer experiments reveal that, compared to the current state-of-the-art quantum methods, QUSL further improves the accuracy and robustness of the task. QUSL achieves up to a $19.5\%$ increase in correlation and reduces the use of critical quantum resources by over $50\%$. As a model framework, QUSL demonstrates transferability applicable to other task scenarios. This work contributes to further integration of classical unsupervised learning techniques with quantum computing, showcasing the potential of quantum methods in complex image processing tasks.

\section*{Authors' contributions}
\textbf{Lian-Hui Yu}: Methodology, Software. \textbf{Xiao-Yu Li}: Conceptualization, Project administration. \textbf{Geng Chen}: Writing - Original Draft, Visualization. \textbf{Qin-Sheng Zhu}: Validation. \textbf{Hui Li}: Formal analysis. \textbf{Guo-Wu Yang}: Supervision.

\section*{Availability of supporting data}
All data generated or analysed during this study are available and included in this published article. \href{https://github.com/QUSL0414/QUSL/tree/main}{Code} for our work has been open-sourced.

\section*{Acknowledgements}
This work was supported by National Key $\mathrm{R\&D}$ Program of China (Grant No.2018FYA0306703), the Open Fund of Advanced Cryptography and System Security Key Laboratory of Sichuan Province (Grant No. SKLACSS-202105), Chengdu Innovation and Technology Project  (No.2021-YF05-02413-GX and 2021-YF09-00114-GX), Sichuan Province key research and development project (No.2022YFG0315).

\small
\bibliographystyle{elsarticle-harv}
\bibliography{quslref}

\begin{thebibliography}{55}
\expandafter\ifx\csname natexlab\endcsname\relax\def\natexlab#1{#1}\fi
\providecommand{\url}[1]{\texttt{#1}}
\providecommand{\href}[2]{#2}
\providecommand{\path}[1]{#1}
\providecommand{\DOIprefix}{doi:}
\providecommand{\ArXivprefix}{arXiv:}
\providecommand{\URLprefix}{URL: }
\providecommand{\Pubmedprefix}{pmid:}
\providecommand{\doi}[1]{\href{http://dx.doi.org/#1}{\path{#1}}}
\providecommand{\Pubmed}[1]{\href{pmid:#1}{\path{#1}}}
\providecommand{\bibinfo}[2]{#2}
\ifx\xfnm\relax \def\xfnm[#1]{\unskip,\space#1}\fi
\bibitem[{Altares-L{\'o}pez et~al.(2021)Altares-L{\'o}pez, Ribeiro and
  Garc{\'\i}a-Ripoll}]{AltaresLópezS}
\bibinfo{author}{Altares-L{\'o}pez, S.}, \bibinfo{author}{Ribeiro, A.},
  \bibinfo{author}{Garc{\'\i}a-Ripoll, J.J.}, \bibinfo{year}{2021}.
\newblock \bibinfo{title}{Automatic design of quantum feature maps}.
\newblock \bibinfo{journal}{Quantum Science and Technology}
  \bibinfo{volume}{6}, \bibinfo{pages}{045015}.
\newblock \DOIprefix\doi{https://doi.org/10.1088/2058-9565/ac1ab1}.
\bibitem[{Arufe et~al.(2022)Arufe, Gonz{\'a}lez, Oddi, Rasconi and
  Varela}]{ArufeLQuantumcircuit}
\bibinfo{author}{Arufe, L.}, \bibinfo{author}{Gonz{\'a}lez, M.A.},
  \bibinfo{author}{Oddi, A.}, \bibinfo{author}{Rasconi, R.},
  \bibinfo{author}{Varela, R.}, \bibinfo{year}{2022}.
\newblock \bibinfo{title}{Quantum circuit compilation by genetic algorithm for
  quantum approximate optimization algorithm applied to maxcut problem}.
\newblock \bibinfo{journal}{Swarm and Evolutionary Computation}
  \bibinfo{volume}{69}, \bibinfo{pages}{101030}.
\newblock \DOIprefix\doi{https://doi.org/10.1016/j.swevo.2022.101030}.
\bibitem[{Arufe et~al.(2023)Arufe, Rasconi, Oddi, Varela and
  Gonz{\'a}lez}]{ArufeLNewcoding}
\bibinfo{author}{Arufe, L.}, \bibinfo{author}{Rasconi, R.},
  \bibinfo{author}{Oddi, A.}, \bibinfo{author}{Varela, R.},
  \bibinfo{author}{Gonz{\'a}lez, M.A.}, \bibinfo{year}{2023}.
\newblock \bibinfo{title}{New coding scheme to compile circuits for quantum
  approximate optimization algorithm by genetic evolution}.
\newblock \bibinfo{journal}{Applied Soft Computing} \bibinfo{volume}{144},
  \bibinfo{pages}{110456}.
\newblock \DOIprefix\doi{https://doi.org/10.1016/j.asoc.2023.110456}.
\bibitem[{Arute et~al.(2019)Arute, Arya, Babbush, Bacon, Bardin, Barends,
  Biswas, Boixo, Brandao, Buell et~al.}]{Arute}
\bibinfo{author}{Arute, F.}, \bibinfo{author}{Arya, K.},
  \bibinfo{author}{Babbush, R.}, \bibinfo{author}{Bacon, D.},
  \bibinfo{author}{Bardin, J.C.}, \bibinfo{author}{Barends, R.},
  \bibinfo{author}{Biswas, R.}, \bibinfo{author}{Boixo, S.},
  \bibinfo{author}{Brandao, F.G.}, \bibinfo{author}{Buell, D.A.}, et~al.,
  \bibinfo{year}{2019}.
\newblock \bibinfo{title}{Quantum supremacy using a programmable
  superconducting processor}.
\newblock \bibinfo{journal}{Nature} \bibinfo{volume}{574},
  \bibinfo{pages}{505--510}.
\newblock \DOIprefix\doi{https://doi.org/10.1038/s41586-019-1666-5}.
\bibitem[{Bai et~al.((2019))Bai, Ding, Bian, Chen, Sun and Wang}]{BaiY}
\bibinfo{author}{Bai, Y.}, \bibinfo{author}{Ding, H.}, \bibinfo{author}{Bian,
  S.}, \bibinfo{author}{Chen, T.}, \bibinfo{author}{Sun, Y.},
  \bibinfo{author}{Wang, W.}, \bibinfo{year}{(2019)}.
\newblock \bibinfo{title}{Simgnn: A neural network approach to fast graph
  similarity computation}, in: \bibinfo{booktitle}{Proceedings of the twelfth
  ACM international conference on web search and data mining}, pp.
  \bibinfo{pages}{384--392}.
\newblock \DOIprefix\doi{https://doi.org/10.1145/3289600.3290967}.
\bibitem[{Benedetti et~al.(2019)Benedetti, Lloyd, Sack and
  Fiorentini}]{BenedettiM}
\bibinfo{author}{Benedetti, M.}, \bibinfo{author}{Lloyd, E.},
  \bibinfo{author}{Sack, S.}, \bibinfo{author}{Fiorentini, M.},
  \bibinfo{year}{2019}.
\newblock \bibinfo{title}{Parameterized quantum circuits as machine learning
  models}.
\newblock \bibinfo{journal}{Quantum Science and Technology}
  \bibinfo{volume}{4}, \bibinfo{pages}{043001}.
\newblock \DOIprefix\doi{https://doi.org/10.1088/2058-9565/ab4eb5}.
\bibitem[{Biamonte et~al.(2017)Biamonte, Wittek, Pancotti, Rebentrost, Wiebe
  and Lloyd}]{Biamonte}
\bibinfo{author}{Biamonte, J.}, \bibinfo{author}{Wittek, P.},
  \bibinfo{author}{Pancotti, N.}, \bibinfo{author}{Rebentrost, P.},
  \bibinfo{author}{Wiebe, N.}, \bibinfo{author}{Lloyd, S.},
  \bibinfo{year}{2017}.
\newblock \bibinfo{title}{Quantum machine learning}.
\newblock \bibinfo{journal}{Nature} \bibinfo{volume}{549},
  \bibinfo{pages}{195--202}.
\newblock \DOIprefix\doi{https://doi.org/10.1038/nature23474}.
\bibitem[{Cerezo et~al.(2021)Cerezo, Arrasmith, Babbush, Benjamin, Endo, Fujii,
  McClean, Mitarai, Yuan, Cincio et~al.}]{CerezoM}
\bibinfo{author}{Cerezo, M.}, \bibinfo{author}{Arrasmith, A.},
  \bibinfo{author}{Babbush, R.}, \bibinfo{author}{Benjamin, S.C.},
  \bibinfo{author}{Endo, S.}, \bibinfo{author}{Fujii, K.},
  \bibinfo{author}{McClean, J.R.}, \bibinfo{author}{Mitarai, K.},
  \bibinfo{author}{Yuan, X.}, \bibinfo{author}{Cincio, L.}, et~al.,
  \bibinfo{year}{2021}.
\newblock \bibinfo{title}{Variational quantum algorithms}.
\newblock \bibinfo{journal}{Nature Reviews Physics} \bibinfo{volume}{3},
  \bibinfo{pages}{625--644}.
\newblock \DOIprefix\doi{https://doi.org/10.1038/s42254-021-00348-9}.
\bibitem[{Chen et~al.(2020)Chen, Kornblith, Norouzi and
  Hinton}]{chen2020simple}
\bibinfo{author}{Chen, T.}, \bibinfo{author}{Kornblith, S.},
  \bibinfo{author}{Norouzi, M.}, \bibinfo{author}{Hinton, G.},
  \bibinfo{year}{2020}.
\newblock \bibinfo{title}{A simple framework for contrastive learning of visual
  representations}, in: \bibinfo{booktitle}{International conference on machine
  learning}, \bibinfo{organization}{PMLR}. pp. \bibinfo{pages}{1597--1607}.
\newblock \URLprefix \url{http://proceedings.mlr.press/v119/chen20j.html}.
\bibitem[{Cheng et~al.(2017)Cheng, Han and Lu}]{ChengG}
\bibinfo{author}{Cheng, G.}, \bibinfo{author}{Han, J.}, \bibinfo{author}{Lu,
  X.}, \bibinfo{year}{2017}.
\newblock \bibinfo{title}{Remote sensing image scene classification: Benchmark
  and state of the art}.
\newblock \bibinfo{journal}{Proceedings of the IEEE} \bibinfo{volume}{105},
  \bibinfo{pages}{1865--1883}.
\newblock \DOIprefix\doi{https://doi.org/10.1109/JPROC.2017.2675998}.
\bibitem[{Cong et~al.(2019)Cong, Choi and Lukin}]{CongI}
\bibinfo{author}{Cong, I.}, \bibinfo{author}{Choi, S.}, \bibinfo{author}{Lukin,
  M.D.}, \bibinfo{year}{2019}.
\newblock \bibinfo{title}{Quantum convolutional neural networks}.
\newblock \bibinfo{journal}{Nature Physics} \bibinfo{volume}{15},
  \bibinfo{pages}{1273--1278}.
\newblock \DOIprefix\doi{https://doi.org/10.1038/s41567-019-0648-8}.
\bibitem[{Dang et~al.(2018)Dang, Jiang, Hu, Ji and Zhang}]{DangY}
\bibinfo{author}{Dang, Y.}, \bibinfo{author}{Jiang, N.}, \bibinfo{author}{Hu,
  H.}, \bibinfo{author}{Ji, Z.}, \bibinfo{author}{Zhang, W.},
  \bibinfo{year}{2018}.
\newblock \bibinfo{title}{Image classification based on quantum
  k-nearest-neighbor algorithm}.
\newblock \bibinfo{journal}{Quantum Information Processing}
  \bibinfo{volume}{17}, \bibinfo{pages}{1--18}.
\newblock \DOIprefix\doi{https://doi.org/10.1007/s11128-018-2004-9}.
\bibitem[{Deng et~al.(2009)Deng, Dong, Socher, Li, Li and
  Fei-Fei}]{deng2009imagenet}
\bibinfo{author}{Deng, J.}, \bibinfo{author}{Dong, W.},
  \bibinfo{author}{Socher, R.}, \bibinfo{author}{Li, L.J.},
  \bibinfo{author}{Li, K.}, \bibinfo{author}{Fei-Fei, L.},
  \bibinfo{year}{2009}.
\newblock \bibinfo{title}{Imagenet: A large-scale hierarchical image database},
  in: \bibinfo{booktitle}{2009 IEEE conference on computer vision and pattern
  recognition}, \bibinfo{organization}{Ieee}. pp. \bibinfo{pages}{248--255}.
\newblock \DOIprefix\doi{https://doi.org/10.1109/CVPR.2009.5206848}.
\bibitem[{Developer(2021)}]{mq_2021}
\bibinfo{author}{Developer, M.}, \bibinfo{year}{2021}.
\newblock \bibinfo{title}{Mindquantum, version 0.9.11}.
\newblock \URLprefix \url{https://gitee.com/mindspore/mindquantum}.
\bibitem[{Ding and Spector((2022))}]{DingL}
\bibinfo{author}{Ding, L.}, \bibinfo{author}{Spector, L.},
  \bibinfo{year}{(2022)}.
\newblock \bibinfo{title}{Evolutionary quantum architecture search for
  parametrized quantum circuits}, in: \bibinfo{booktitle}{Proceedings of the
  Genetic and Evolutionary Computation Conference Companion}, pp.
  \bibinfo{pages}{2190--2195}.
\newblock \DOIprefix\doi{https://doi.org/10.1145/3520304.3534012}.
\bibitem[{Douze et~al.(2021)Douze, Tolias, Pizzi, Papakipos, Chanussot,
  Radenovic, Jenicek, Maximov, Leal-Taix{\'e}, Elezi et~al.}]{DouzeM}
\bibinfo{author}{Douze, M.}, \bibinfo{author}{Tolias, G.},
  \bibinfo{author}{Pizzi, E.}, \bibinfo{author}{Papakipos, Z.},
  \bibinfo{author}{Chanussot, L.}, \bibinfo{author}{Radenovic, F.},
  \bibinfo{author}{Jenicek, T.}, \bibinfo{author}{Maximov, M.},
  \bibinfo{author}{Leal-Taix{\'e}, L.}, \bibinfo{author}{Elezi, I.}, et~al.,
  \bibinfo{year}{2021}.
\newblock \bibinfo{title}{The 2021 image similarity dataset and challenge}.
\newblock \bibinfo{journal}{arXiv preprint arXiv:2106.09672} \URLprefix
  \url{https://arxiv.org/abs/2106.09672}.
\bibitem[{Eisert(2021)}]{EisertJ}
\bibinfo{author}{Eisert, J.}, \bibinfo{year}{2021}.
\newblock \bibinfo{title}{Entangling power and quantum circuit complexity}.
\newblock \bibinfo{journal}{Physical Review Letters} \bibinfo{volume}{127},
  \bibinfo{pages}{020501}.
\newblock \DOIprefix\doi{https://doi.org/10.1103/PhysRevLett.127.020501}.
\bibitem[{Fang et~al.(2008)Fang, Wang, Tu and Horstemeyer}]{fang2008efficient}
\bibinfo{author}{Fang, H.}, \bibinfo{author}{Wang, Q.}, \bibinfo{author}{Tu,
  Y.C.}, \bibinfo{author}{Horstemeyer, M.F.}, \bibinfo{year}{2008}.
\newblock \bibinfo{title}{An efficient non-dominated sorting method for
  evolutionary algorithms}.
\newblock \bibinfo{journal}{Evolutionary computation} \bibinfo{volume}{16},
  \bibinfo{pages}{355--384}.
\newblock \DOIprefix\doi{https://doi.org/10.1162/evco.2008.16.3.355}.
\bibitem[{He et~al.((2020))He, Fan, Wu, Xie and Girshick}]{he2020momentum}
\bibinfo{author}{He, K.}, \bibinfo{author}{Fan, H.}, \bibinfo{author}{Wu, Y.},
  \bibinfo{author}{Xie, S.}, \bibinfo{author}{Girshick, R.},
  \bibinfo{year}{(2020)}.
\newblock \bibinfo{title}{Momentum contrast for unsupervised visual
  representation learning}, in: \bibinfo{booktitle}{Proceedings of the IEEE/CVF
  conference on computer vision and pattern recognition}, pp.
  \bibinfo{pages}{9729--9738}.
\newblock \URLprefix
  \url{https://openaccess.thecvf.com/content_CVPR_2020/html/He_Momentum_Contrast_for_Unsupervised_Visual_Representation_Learning_CVPR_2020_paper.html}.
\bibitem[{Hendrycks and Dietterich(2019)}]{Hendrycks}
\bibinfo{author}{Hendrycks, D.}, \bibinfo{author}{Dietterich, T.},
  \bibinfo{year}{2019}.
\newblock \bibinfo{title}{Benchmarking neural network robustness to common
  corruptions and perturbations}.
\newblock \bibinfo{journal}{arXiv preprint arXiv:1903.12261} \URLprefix
  \url{https://arxiv.org/abs/1903.12261}.
\bibitem[{Huang et~al.(2022)Huang, Broughton, Cotler, Chen, Li, Mohseni, Neven,
  Babbush, Kueng, Preskill et~al.}]{HuangHYQuantumadvantage}
\bibinfo{author}{Huang, H.Y.}, \bibinfo{author}{Broughton, M.},
  \bibinfo{author}{Cotler, J.}, \bibinfo{author}{Chen, S.},
  \bibinfo{author}{Li, J.}, \bibinfo{author}{Mohseni, M.},
  \bibinfo{author}{Neven, H.}, \bibinfo{author}{Babbush, R.},
  \bibinfo{author}{Kueng, R.}, \bibinfo{author}{Preskill, J.}, et~al.,
  \bibinfo{year}{2022}.
\newblock \bibinfo{title}{Quantum advantage in learning from experiments}.
\newblock \bibinfo{journal}{Science} \bibinfo{volume}{376},
  \bibinfo{pages}{1182--1186}.
\newblock \DOIprefix\doi{https://doi.org/10.1126/science.abn7293}.
\bibitem[{Huang et~al.(2021)Huang, Broughton, Mohseni, Babbush, Boixo, Neven
  and McClean}]{HuangB}
\bibinfo{author}{Huang, H.Y.}, \bibinfo{author}{Broughton, M.},
  \bibinfo{author}{Mohseni, M.}, \bibinfo{author}{Babbush, R.},
  \bibinfo{author}{Boixo, S.}, \bibinfo{author}{Neven, H.},
  \bibinfo{author}{McClean, J.R.}, \bibinfo{year}{2021}.
\newblock \bibinfo{title}{Power of data in quantum machine learning}.
\newblock \bibinfo{journal}{Nature communications} \bibinfo{volume}{12},
  \bibinfo{pages}{2631}.
\newblock \DOIprefix\doi{https://doi.org/10.1038/s41467-021-22539-9}.
\bibitem[{Jaderberg et~al.(2022)Jaderberg, Anderson, Xie, Albanie, Kiffner and
  Jaksch}]{jaderberg2022quantum}
\bibinfo{author}{Jaderberg, B.}, \bibinfo{author}{Anderson, L.W.},
  \bibinfo{author}{Xie, W.}, \bibinfo{author}{Albanie, S.},
  \bibinfo{author}{Kiffner, M.}, \bibinfo{author}{Jaksch, D.},
  \bibinfo{year}{2022}.
\newblock \bibinfo{title}{Quantum self-supervised learning}.
\newblock \bibinfo{journal}{Quantum Science and Technology}
  \bibinfo{volume}{7}, \bibinfo{pages}{035005}.
\newblock \DOIprefix\doi{https://doi.org/10.1088/2058-9565/ac6825}.
\bibitem[{Knill(2010)}]{Knill}
\bibinfo{author}{Knill, E.}, \bibinfo{year}{2010}.
\newblock \bibinfo{title}{Quantum computing}.
\newblock \bibinfo{journal}{Nature} \bibinfo{volume}{463},
  \bibinfo{pages}{441--443}.
\newblock \DOIprefix\doi{https://doi.org/10.1038/463441a}.
\bibitem[{Krizhevsky et~al.(2009)Krizhevsky, Hinton
  et~al.}]{krizhevsky2009learning}
\bibinfo{author}{Krizhevsky, A.}, \bibinfo{author}{Hinton, G.}, et~al.,
  \bibinfo{year}{2009}.
\newblock \bibinfo{title}{Learning multiple layers of features from tiny
  images} \DOIprefix\doi{https://doi.org/10.1609/aaai.v34i07.6845}.
\bibitem[{Krylov and Lukac((2019))}]{KrylovG}
\bibinfo{author}{Krylov, G.}, \bibinfo{author}{Lukac, M.},
  \bibinfo{year}{(2019)}.
\newblock \bibinfo{title}{Quantum encoded quantum evolutionary algorithm for
  the design of quantum circuits}, in: \bibinfo{booktitle}{Proceedings of the
  16th ACM International Conference on Computing Frontiers}, pp.
  \bibinfo{pages}{220--225}.
\newblock \DOIprefix\doi{https://doi.org/10.1145/3310273.3322826}.
\bibitem[{Li et~al.(2024)Li, Zhu, Hu, Wu, Yang, Yu and Chen}]{li2024new}
\bibinfo{author}{Li, X.Y.}, \bibinfo{author}{Zhu, Q.S.}, \bibinfo{author}{Hu,
  Y.}, \bibinfo{author}{Wu, H.}, \bibinfo{author}{Yang, G.W.},
  \bibinfo{author}{Yu, L.H.}, \bibinfo{author}{Chen, G.}, \bibinfo{year}{2024}.
\newblock \bibinfo{title}{A new quantum machine learning algorithm: split
  hidden quantum markov model inspired by quantum conditional master equation}.
\newblock \bibinfo{journal}{Quantum} \bibinfo{volume}{8},
  \bibinfo{pages}{1232}.
\newblock \DOIprefix\doi{https://doi.org/10.22331/q-2024-01-24-1232}.
\bibitem[{Lin et~al.((2014))Lin, Maire, Belongie, Hays, Perona, Ramanan,
  Doll{\'a}r and Zitnick}]{LinTY}
\bibinfo{author}{Lin, T.Y.}, \bibinfo{author}{Maire, M.},
  \bibinfo{author}{Belongie, S.}, \bibinfo{author}{Hays, J.},
  \bibinfo{author}{Perona, P.}, \bibinfo{author}{Ramanan, D.},
  \bibinfo{author}{Doll{\'a}r, P.}, \bibinfo{author}{Zitnick, C.L.},
  \bibinfo{year}{(2014)}.
\newblock \bibinfo{title}{Microsoft coco: Common objects in context}, in:
  \bibinfo{booktitle}{Computer Vision--ECCV 2014: 13th European Conference,
  Zurich, Switzerland, September 6-12, 2014, Proceedings, Part V 13},
  \bibinfo{organization}{Springer}. pp. \bibinfo{pages}{740--755}.
\newblock \DOIprefix\doi{https://doi.org/10.1007/978-3-319-10602-1_48}.
\bibitem[{Liu et~al.(2023)Liu, Li, Duan, Tsang and Yang}]{LiuD}
\bibinfo{author}{Liu, D.}, \bibinfo{author}{Li, W.}, \bibinfo{author}{Duan,
  L.}, \bibinfo{author}{Tsang, I.W.}, \bibinfo{author}{Yang, G.},
  \bibinfo{year}{2023}.
\newblock \bibinfo{title}{Noisy label learning with provable consistency for a
  wider family of losses}.
\newblock \bibinfo{journal}{IEEE Transactions on Pattern Analysis and Machine
  Intelligence} \DOIprefix\doi{https://doi.org/10.1109/TPAMI.2023.3296156}.
\bibitem[{Liu et~al.(2019)Liu, Zhou, El-Rafei, Li and Xu}]{LiuXA}
\bibinfo{author}{Liu, X.}, \bibinfo{author}{Zhou, R.G.},
  \bibinfo{author}{El-Rafei, A.}, \bibinfo{author}{Li, F.X.},
  \bibinfo{author}{Xu, R.Q.}, \bibinfo{year}{2019}.
\newblock \bibinfo{title}{Similarity assessment of quantum images}.
\newblock \bibinfo{journal}{Quantum Information Processing}
  \bibinfo{volume}{18}, \bibinfo{pages}{1--19}.
\newblock \DOIprefix\doi{https://doi.org/10.1007/s11128-019-2357-8}.
\bibitem[{Ma et~al.(2020)Ma, Dong, Long, Zhang, He, Xue and Ji}]{ma2020fine}
\bibinfo{author}{Ma, Z.}, \bibinfo{author}{Dong, J.}, \bibinfo{author}{Long,
  Z.}, \bibinfo{author}{Zhang, Y.}, \bibinfo{author}{He, Y.},
  \bibinfo{author}{Xue, H.}, \bibinfo{author}{Ji, S.}, \bibinfo{year}{2020}.
\newblock \bibinfo{title}{Fine-grained fashion similarity learning by
  attribute-specific embedding network}, in: \bibinfo{booktitle}{Proceedings of
  the AAAI Conference on artificial intelligence}, pp.
  \bibinfo{pages}{11741--11748}.
\newblock \DOIprefix\doi{https://doi.org/10.1609/aaai.v34i07.6845}.
\bibitem[{Mei et~al.(2023)Mei, Fan, Zhang, Yu, Zhou, Liu, Fu, Huang and
  Shi}]{MeiY}
\bibinfo{author}{Mei, Y.}, \bibinfo{author}{Fan, Y.}, \bibinfo{author}{Zhang,
  Y.}, \bibinfo{author}{Yu, J.}, \bibinfo{author}{Zhou, Y.},
  \bibinfo{author}{Liu, D.}, \bibinfo{author}{Fu, Y.}, \bibinfo{author}{Huang,
  T.S.}, \bibinfo{author}{Shi, H.}, \bibinfo{year}{2023}.
\newblock \bibinfo{title}{Pyramid attention network for image restoration}.
\newblock \bibinfo{journal}{International Journal of Computer Vision}
  \bibinfo{volume}{131}, \bibinfo{pages}{3207--3225}.
\newblock \DOIprefix\doi{https://doi.org/10.1007/s11263-023-01843-5}.
\bibitem[{Men et~al.((2022))Men, Yao, Cui, Lian, Xie and Hua}]{MenY}
\bibinfo{author}{Men, Y.}, \bibinfo{author}{Yao, Y.}, \bibinfo{author}{Cui,
  M.}, \bibinfo{author}{Lian, Z.}, \bibinfo{author}{Xie, X.},
  \bibinfo{author}{Hua, X.S.}, \bibinfo{year}{(2022)}.
\newblock \bibinfo{title}{Unpaired cartoon image synthesis via gated cycle
  mapping}, in: \bibinfo{booktitle}{Proceedings of the IEEE/CVF conference on
  computer vision and pattern recognition}, pp. \bibinfo{pages}{3501--3510}.
\newblock \URLprefix
  \url{https://openaccess.thecvf.com/content/CVPR2022/html/Men_Unpaired_Cartoon_Image_Synthesis_via_Gated_Cycle_Mapping_CVPR_2022_paper.html}.
\bibitem[{Mohseni et~al.(2022)Mohseni, McMahon and Byrnes}]{MohseniN}
\bibinfo{author}{Mohseni, N.}, \bibinfo{author}{McMahon, P.L.},
  \bibinfo{author}{Byrnes, T.}, \bibinfo{year}{2022}.
\newblock \bibinfo{title}{Ising machines as hardware solvers of combinatorial
  optimization problems}.
\newblock \bibinfo{journal}{Nature Reviews Physics} \bibinfo{volume}{4},
  \bibinfo{pages}{363--379}.
\newblock \DOIprefix\doi{https://doi.org/10.1038/s42254-022-00440-8}.
\bibitem[{Nam et~al.(2018)Nam, Ross, Su, Childs and Maslov}]{NamY}
\bibinfo{author}{Nam, Y.}, \bibinfo{author}{Ross, N.J.}, \bibinfo{author}{Su,
  Y.}, \bibinfo{author}{Childs, A.M.}, \bibinfo{author}{Maslov, D.},
  \bibinfo{year}{2018}.
\newblock \bibinfo{title}{Automated optimization of large quantum circuits with
  continuous parameters}.
\newblock \bibinfo{journal}{npj Quantum Information} \bibinfo{volume}{4},
  \bibinfo{pages}{23}.
\newblock \DOIprefix\doi{https://doi.org/10.1038/s41534-018-0072-4}.
\bibitem[{Paine et~al.(2023)Paine, Elfving and Kyriienko}]{PaineAE}
\bibinfo{author}{Paine, A.E.}, \bibinfo{author}{Elfving, V.E.},
  \bibinfo{author}{Kyriienko, O.}, \bibinfo{year}{2023}.
\newblock \bibinfo{title}{Quantum kernel methods for solving regression
  problems and differential equations}.
\newblock \bibinfo{journal}{Physical Review A} \bibinfo{volume}{107},
  \bibinfo{pages}{032428}.
\newblock \DOIprefix\doi{https://doi.org/10.1103/PhysRevA.107.032428}.
\bibitem[{Palubinskas(2017)}]{PalubinskasG}
\bibinfo{author}{Palubinskas, G.}, \bibinfo{year}{2017}.
\newblock \bibinfo{title}{Image similarity/distance measures: what is really
  behind mse and ssim?}
\newblock \bibinfo{journal}{International Journal of Image and Data Fusion}
  \bibinfo{volume}{8}, \bibinfo{pages}{32--53}.
\newblock \DOIprefix\doi{https://doi.org/10.1080/19479832.2016.1273259}.
\bibitem[{P{\'e}rez-Salinas et~al.(2020)P{\'e}rez-Salinas, Cervera-Lierta,
  Gil-Fuster and Latorre}]{PérezSalinasA}
\bibinfo{author}{P{\'e}rez-Salinas, A.}, \bibinfo{author}{Cervera-Lierta, A.},
  \bibinfo{author}{Gil-Fuster, E.}, \bibinfo{author}{Latorre, J.I.},
  \bibinfo{year}{2020}.
\newblock \bibinfo{title}{Data re-uploading for a universal quantum
  classifier}.
\newblock \bibinfo{journal}{Quantum} \bibinfo{volume}{4}, \bibinfo{pages}{226}.
\newblock \DOIprefix\doi{https://doi.org/10.22331/q-2020-02-06-226}.
\bibitem[{Preskill(2018)}]{Preskill}
\bibinfo{author}{Preskill, J.}, \bibinfo{year}{2018}.
\newblock \bibinfo{title}{Quantum computing in the nisq era and beyond}.
\newblock \bibinfo{journal}{Quantum} \bibinfo{volume}{2}, \bibinfo{pages}{79}.
\newblock \DOIprefix\doi{https://doi.org/10.22331/q-2018-08-06-79}.
\bibitem[{Rasconi and Oddi((2019))}]{RasconiR}
\bibinfo{author}{Rasconi, R.}, \bibinfo{author}{Oddi, A.},
  \bibinfo{year}{(2019)}.
\newblock \bibinfo{title}{An innovative genetic algorithm for the quantum
  circuit compilation problem}, in: \bibinfo{booktitle}{Proceedings of the AAAI
  conference on artificial intelligence}, pp. \bibinfo{pages}{7707--7714}.
\newblock \DOIprefix\doi{https://doi.org/10.1609/aaai.v33i01.33017707}.
\bibitem[{Rebentrost et~al.(2014)Rebentrost, Mohseni and Lloyd}]{Rebentrost}
\bibinfo{author}{Rebentrost, P.}, \bibinfo{author}{Mohseni, M.},
  \bibinfo{author}{Lloyd, S.}, \bibinfo{year}{2014}.
\newblock \bibinfo{title}{Quantum support vector machine for big data
  classification}.
\newblock \bibinfo{journal}{Physical review letters} \bibinfo{volume}{113},
  \bibinfo{pages}{130503}.
\newblock \DOIprefix\doi{https://doi.org/10.1103/PhysRevLett.113.130503}.
\bibitem[{Schuld et~al.(2014)Schuld, Sinayskiy and Petruccione}]{SchuldM}
\bibinfo{author}{Schuld, M.}, \bibinfo{author}{Sinayskiy, I.},
  \bibinfo{author}{Petruccione, F.}, \bibinfo{year}{2014}.
\newblock \bibinfo{title}{The quest for a quantum neural network}.
\newblock \bibinfo{journal}{Quantum Information Processing}
  \bibinfo{volume}{13}, \bibinfo{pages}{2567--2586}.
\newblock \DOIprefix\doi{https://doi.org/10.1007/s11128-014-0809-8}.
\bibitem[{Schuld et~al.(2015)Schuld, Sinayskiy and Petruccione}]{Schuld}
\bibinfo{author}{Schuld, M.}, \bibinfo{author}{Sinayskiy, I.},
  \bibinfo{author}{Petruccione, F.}, \bibinfo{year}{2015}.
\newblock \bibinfo{title}{An introduction to quantum machine learning}.
\newblock \bibinfo{journal}{Contemporary Physics} \bibinfo{volume}{56},
  \bibinfo{pages}{172--185}.
\newblock \DOIprefix\doi{https://doi.org/10.1080/00107514.2014.964942}.
\bibitem[{Silver et~al.((2023))Silver, Patel, Ranjan, Gandhi, Cutler and
  Tiwari}]{SilverDSLIQ}
\bibinfo{author}{Silver, D.}, \bibinfo{author}{Patel, T.},
  \bibinfo{author}{Ranjan, A.}, \bibinfo{author}{Gandhi, H.},
  \bibinfo{author}{Cutler, W.}, \bibinfo{author}{Tiwari, D.},
  \bibinfo{year}{(2023)}.
\newblock \bibinfo{title}{Sliq: quantum image similarity networks on noisy
  quantum computers}, in: \bibinfo{booktitle}{Proceedings of the AAAI
  Conference on Artificial Intelligence}, pp. \bibinfo{pages}{9846--9854}.
\newblock \DOIprefix\doi{https://doi.org/10.1609/aaai.v37i8.26175}.
\bibitem[{Sim et~al.(2019)Sim, Johnson and
  Aspuru-Guzik}]{sim2019expressibility}
\bibinfo{author}{Sim, S.}, \bibinfo{author}{Johnson, P.D.},
  \bibinfo{author}{Aspuru-Guzik, A.}, \bibinfo{year}{2019}.
\newblock \bibinfo{title}{Expressibility and entangling capability of
  parameterized quantum circuits for hybrid quantum-classical algorithms}.
\newblock \bibinfo{journal}{Advanced Quantum Technologies} \bibinfo{volume}{2},
  \bibinfo{pages}{1900070}.
\newblock \DOIprefix\doi{https://doi.org/10.1002/qute.201900070}.
\bibitem[{Veit et~al.((2017))Veit, Belongie and Karaletsos}]{VeitA}
\bibinfo{author}{Veit, A.}, \bibinfo{author}{Belongie, S.},
  \bibinfo{author}{Karaletsos, T.}, \bibinfo{year}{(2017)}.
\newblock \bibinfo{title}{Conditional similarity networks}, in:
  \bibinfo{booktitle}{Proceedings of the IEEE conference on computer vision and
  pattern recognition}, pp. \bibinfo{pages}{830--838}.
\newblock \URLprefix
  \url{https://openaccess.thecvf.com/content_cvpr_2017/html/Veit_Conditional_Similarity_Networks_CVPR_2017_paper.html}.
\bibitem[{Wang et~al.(2022)Wang, Ding, Gu, Lin, Pan, Chong and Han}]{WangH}
\bibinfo{author}{Wang, H.}, \bibinfo{author}{Ding, Y.}, \bibinfo{author}{Gu,
  J.}, \bibinfo{author}{Lin, Y.}, \bibinfo{author}{Pan, D.Z.},
  \bibinfo{author}{Chong, F.T.}, \bibinfo{author}{Han, S.},
  \bibinfo{year}{2022}.
\newblock \bibinfo{title}{Quantumnas: Noise-adaptive search for robust quantum
  circuits}, in: \bibinfo{booktitle}{2022 IEEE International Symposium on
  High-Performance Computer Architecture (HPCA)}, \bibinfo{organization}{IEEE}.
  pp. \bibinfo{pages}{692--708}.
\newblock \DOIprefix\doi{https://doi.org/10.1109/HPCA53966.2022.00057}.
\bibitem[{Wang et~al.((2014))Wang, Song, Leung, Rosenberg, Wang, Philbin, Chen
  and Wu}]{WangJ}
\bibinfo{author}{Wang, J.}, \bibinfo{author}{Song, Y.}, \bibinfo{author}{Leung,
  T.}, \bibinfo{author}{Rosenberg, C.}, \bibinfo{author}{Wang, J.},
  \bibinfo{author}{Philbin, J.}, \bibinfo{author}{Chen, B.},
  \bibinfo{author}{Wu, Y.}, \bibinfo{year}{(2014)}.
\newblock \bibinfo{title}{Learning fine-grained image similarity with deep
  ranking}, in: \bibinfo{booktitle}{Proceedings of the IEEE conference on
  computer vision and pattern recognition}, pp. \bibinfo{pages}{1386--1393}.
\newblock \URLprefix
  \url{https://openaccess.thecvf.com/content_cvpr_2014/html/Wang_Learning_Fine-grained_Image_2014_CVPR_paper.html}.
\bibitem[{Wang et~al.(2023)Wang, Du, Yang, Zhang, Wang, Zhang, Yang, Huang and
  Han}]{WangX}
\bibinfo{author}{Wang, X.}, \bibinfo{author}{Du, Y.}, \bibinfo{author}{Yang,
  S.}, \bibinfo{author}{Zhang, J.}, \bibinfo{author}{Wang, M.},
  \bibinfo{author}{Zhang, J.}, \bibinfo{author}{Yang, W.},
  \bibinfo{author}{Huang, J.}, \bibinfo{author}{Han, X.}, \bibinfo{year}{2023}.
\newblock \bibinfo{title}{Retccl: Clustering-guided contrastive learning for
  whole-slide image retrieval}.
\newblock \bibinfo{journal}{Medical image analysis} \bibinfo{volume}{83},
  \bibinfo{pages}{102645}.
\newblock \DOIprefix\doi{https://doi.org/10.1016/j.media.2022.102645}.
\bibitem[{Wu et~al.((2023))Wu, Yan, Lu, Pan and Yan}]{WuW}
\bibinfo{author}{Wu, W.}, \bibinfo{author}{Yan, G.}, \bibinfo{author}{Lu, X.},
  \bibinfo{author}{Pan, K.}, \bibinfo{author}{Yan, J.}, \bibinfo{year}{(2023)}.
\newblock \bibinfo{title}{Quantumdarts: differentiable quantum architecture
  search for variational quantum algorithms}, in:
  \bibinfo{booktitle}{International Conference on Machine Learning},
  \bibinfo{organization}{PMLR}. pp. \bibinfo{pages}{37745--37764}.
\newblock \URLprefix \url{https://proceedings.mlr.press/v202/wu23v.html}.
\bibitem[{Xu et~al.(2021)Xu, Liu, Cao, Huang, Liu, Qian, Liu, Wu, Dong, Qiu
  et~al.}]{XuY}
\bibinfo{author}{Xu, Y.}, \bibinfo{author}{Liu, X.}, \bibinfo{author}{Cao, X.},
  \bibinfo{author}{Huang, C.}, \bibinfo{author}{Liu, E.},
  \bibinfo{author}{Qian, S.}, \bibinfo{author}{Liu, X.}, \bibinfo{author}{Wu,
  Y.}, \bibinfo{author}{Dong, F.}, \bibinfo{author}{Qiu, C.W.}, et~al.,
  \bibinfo{year}{2021}.
\newblock \bibinfo{title}{Artificial intelligence: A powerful paradigm for
  scientific research}.
\newblock \bibinfo{journal}{The Innovation} \bibinfo{volume}{2}.
\newblock \DOIprefix\doi{https://doi.org/10.1016/j.xinn.2021.100179}.
\bibitem[{Yan et~al.((2012))Yan, Le, Iliyasu, Sun, Garcia, Dong and
  Hirota}]{YanF}
\bibinfo{author}{Yan, F.}, \bibinfo{author}{Le, P.Q.},
  \bibinfo{author}{Iliyasu, A.M.}, \bibinfo{author}{Sun, B.},
  \bibinfo{author}{Garcia, J.A.}, \bibinfo{author}{Dong, F.},
  \bibinfo{author}{Hirota, K.}, \bibinfo{year}{(2012)}.
\newblock \bibinfo{title}{Assessing the similarity of quantum images based on
  probability measurements}, in: \bibinfo{booktitle}{2012 IEEE Congress on
  Evolutionary Computation}, \bibinfo{organization}{IEEE}. pp.
  \bibinfo{pages}{1--6}.
\newblock \DOIprefix\doi{https://doi.org/10.1109/CEC.2012.6256418}.
\bibitem[{Zhang and Zhao(2023)}]{ZhangA}
\bibinfo{author}{Zhang, A.}, \bibinfo{author}{Zhao, S.}, \bibinfo{year}{2023}.
\newblock \bibinfo{title}{Evolutionary-based searching method for quantum
  circuit architecture}.
\newblock \bibinfo{journal}{Quantum Information Processing}
  \bibinfo{volume}{22}, \bibinfo{pages}{283}.
\newblock \DOIprefix\doi{https://doi.org/10.1007/s11128-023-04033-x}.
\bibitem[{Zhang et~al.(2011)Zhang, Zhang, Mou and Zhang}]{ZhangL}
\bibinfo{author}{Zhang, L.}, \bibinfo{author}{Zhang, L.}, \bibinfo{author}{Mou,
  X.}, \bibinfo{author}{Zhang, D.}, \bibinfo{year}{2011}.
\newblock \bibinfo{title}{Fsim: A feature similarity index for image quality
  assessment}.
\newblock \bibinfo{journal}{IEEE transactions on Image Processing}
  \bibinfo{volume}{20}, \bibinfo{pages}{2378--2386}.
\newblock \DOIprefix\doi{https://doi.org/10.1109/TIP.2011.2109730}.
\bibitem[{Zhou and Sun(2015)}]{ZhouRG}
\bibinfo{author}{Zhou, R.G.}, \bibinfo{author}{Sun, Y.J.},
  \bibinfo{year}{2015}.
\newblock \bibinfo{title}{Quantum multidimensional color images similarity
  comparison}.
\newblock \bibinfo{journal}{Quantum Information Processing}
  \bibinfo{volume}{14}, \bibinfo{pages}{1605--1624}.
\newblock \DOIprefix\doi{https://doi.org/10.1007/s11128-014-0849-0}.

\end{thebibliography}

\newpage

\appendix 

\section{The quantum circuits obtained by QUSL}
\label{appendixA}

The quantum circuits with the best expressive capabilities obtained by QUSL on landscape and DISC21 are shown below. Summarizing the optimal circuits obtained across all datasets reveals some interesting patterns and structures. $\{R_X, R_Y, R_Z\}$ and $CNOT$ gates present complex interweaving, forming an intricate high-dimensional Hilbert space. The $CNOT$-$R_X$-$R_Y$-$R_Z$ sequence appears frequently in multiple circuits, potentially serving as an effective module. Another prominent pattern is multiple qubits connected to a single qubit through $CNOT$ gates, which may effectively transform this qubit into an attention center through entanglement, thereby achieving crucial feature extraction.

\begin{figure}[H]
	\centering
	\label{AF1}
	\includegraphics[width=0.78\textwidth, trim=1.5cm 1cm 0.5cm 1cm,clip]{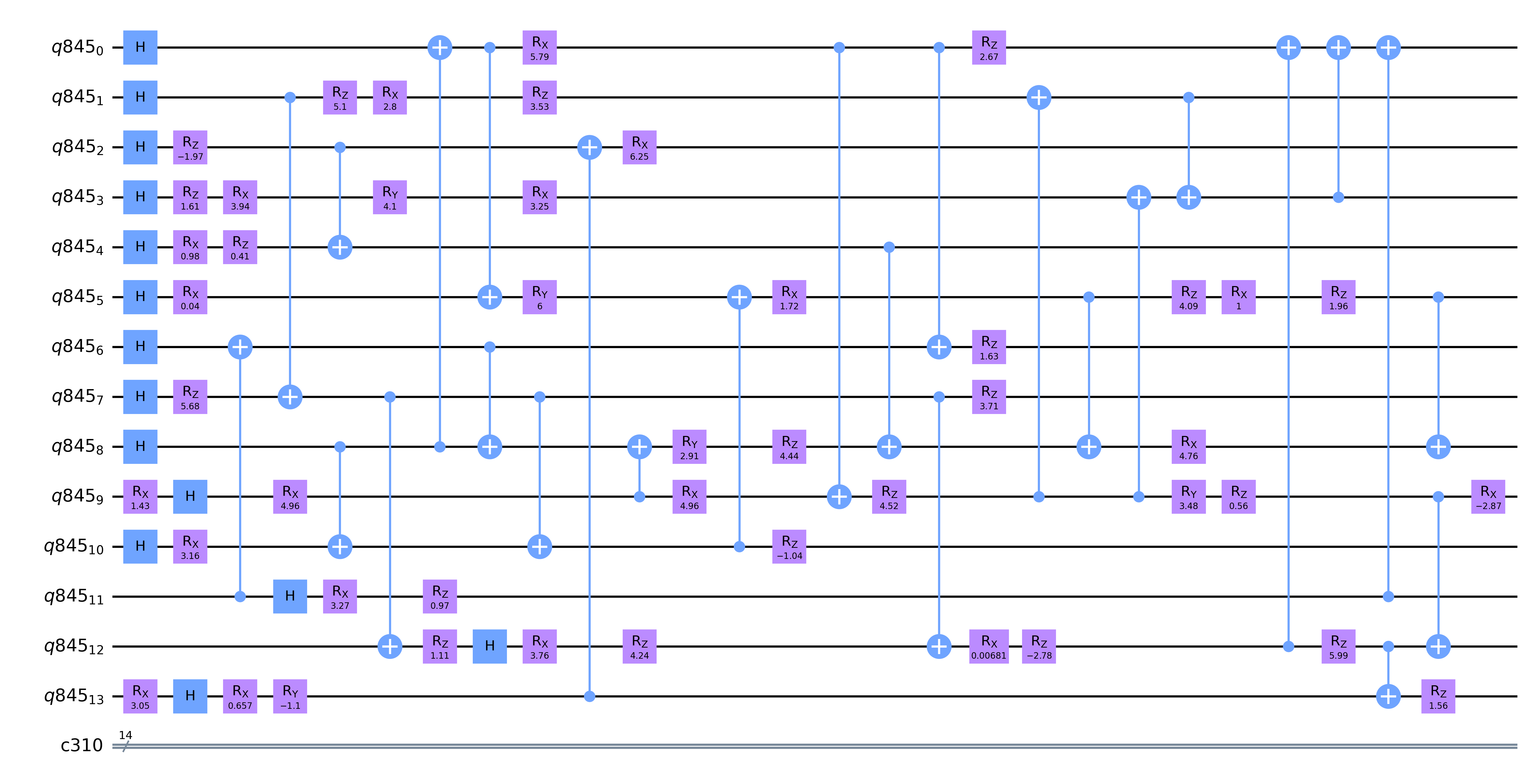}
	\caption{
		\footnotesize
		The optimal-performing quantum circuit obtained by QUSL on the landscape, with $26$ CNOT operations and a circuit depth of $28$.}
\end{figure}

\begin{figure}[H]
	\centering
	\label{AF2}
	\includegraphics[width=0.78\textwidth, trim=1.5cm 1cm 0.5cm 1cm,clip]{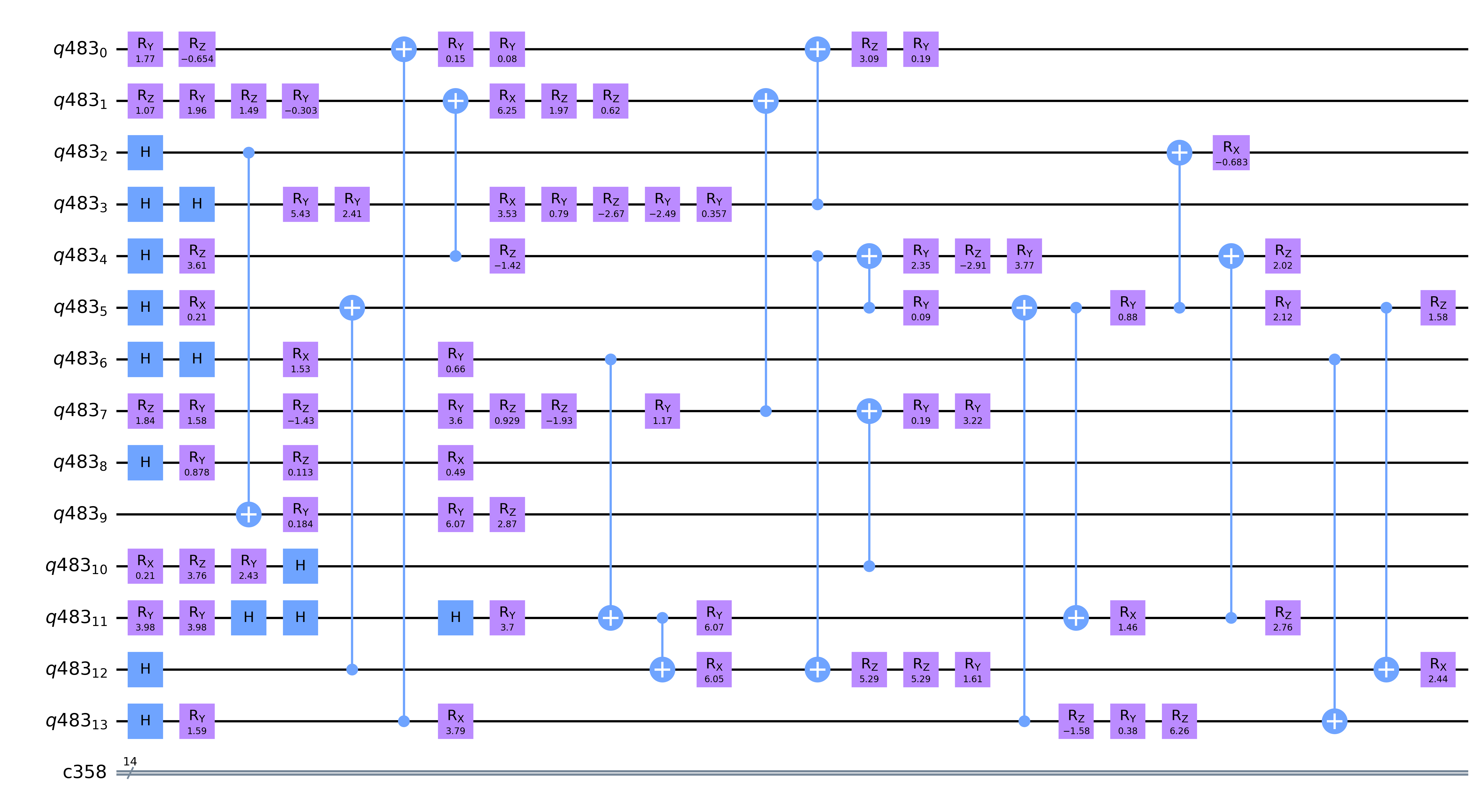}
	\caption{
		\footnotesize
		The optimal-performing quantum circuit obtained by QUSL on the DISC21, with $17$ CNOT operations and a circuit depth of $26$.}
\end{figure}

\section{Quantum circuits with similar fitness but significantly different structures}
\label{appendixB}

During the training process, some circuits with similar expressive capabilities but enormous structural differences emerged, as shown in the figure below. These circuits are derived from the training process on the COCO dataset, exhibiting fitness differences on the order of $10^{-4}$, yet their quantum circuit depths and CNOT counts differ by almost a factor of two. This phenomenon poses challenges for population diversity.

\begin{figure}[H]
	\centering
	\includegraphics[width=0.85\textwidth, trim=9cm 0cm 8cm 0cm,clip]{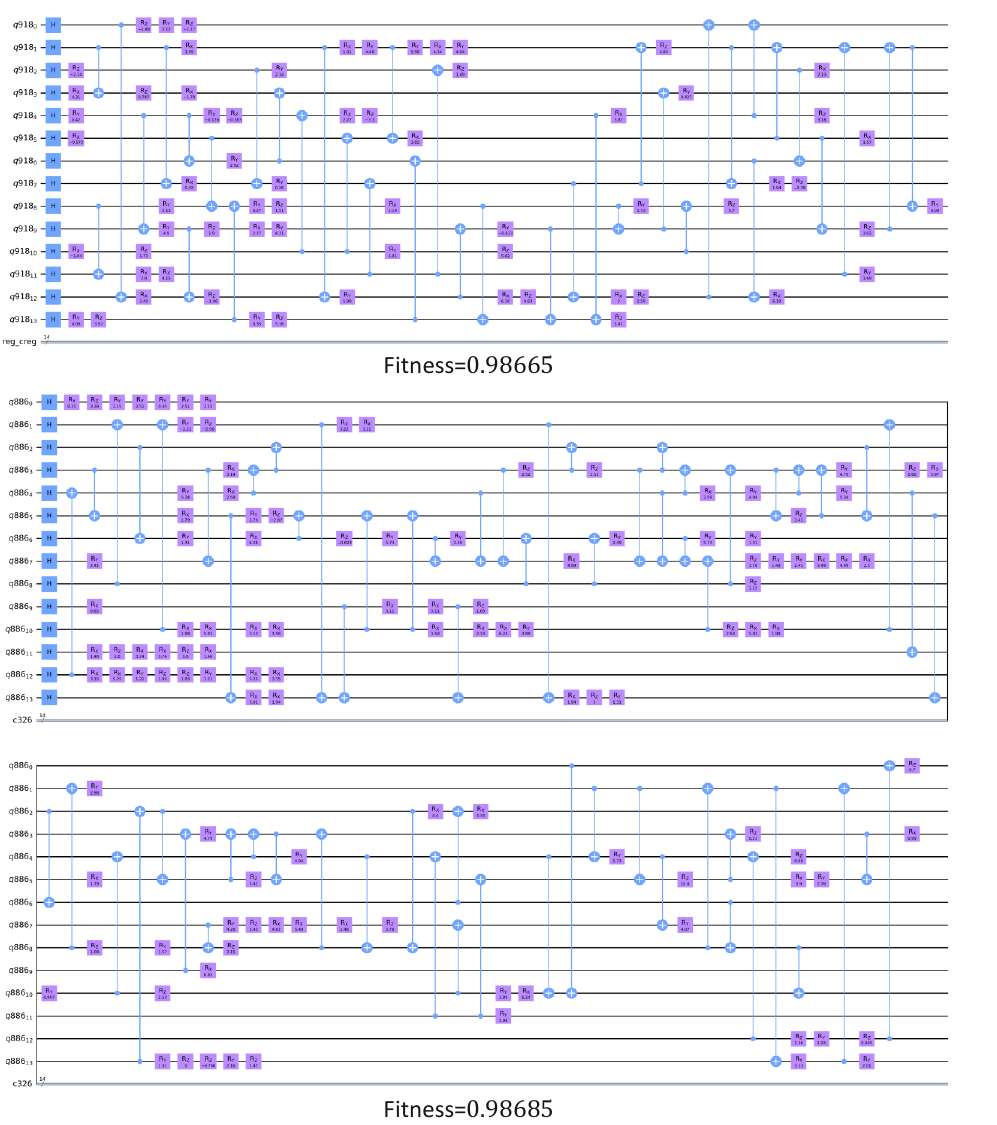}
	\caption{
		\footnotesize
		Quantum circuits with significant structural differences but similar fitness levels.}
\end{figure}

\section{Performance Evaluation of QUSL in classification tasks}
\label{appendixC}

We extend our evaluation to include multi-classification tasks as simplified instances of image similarity tasks as explained in the introduction, offering insights into QUSL's performance. Rigorous binary classification experiments on three widely recognized multi-classification datasets — mnist, fashion-mnist, and AIDS — aim to validate the model's usability and assess its performance under simplified scenarios. Table \ref{Classical Table} compares the performance of SliQ and QUSL on three multi-classification datasets.

\begin{table}[htb]
	\label{Classical Table}
	\renewcommand{\arraystretch}{1.2}
	\caption{The comparison between QUSL and SliQ in terms of classification task accuracy}
	\centering
	\scriptsize
	\begin{threeparttable}
		\begin{tabular}{*{3}{>{\centering\arraybackslash}p{2.5cm}}}
			\toprule
			Dataset & SliQ & QUSL (ours) \\[-0.1cm]
			\midrule
			mnist & $97.16(\pm 0.193)$ &$98.50(\pm 0.235$)   \\
			fashion-mnist &  $92.53(\pm 0.409)$&$92.76(\pm0.682)$  \\
			AIDS &  $71.54(\pm0.257)$&$71.32(\pm 0.457)$  \\
			\bottomrule
		\end{tabular}
	\end{threeparttable}
\end{table}

In classification tasks, QUSL and SliQ exhibit similar accuracy, but with slightly reduced robustness in the case of QUSL. This may be attributed to the heuristic algorithm-based quantum circuit design facing more pronounced challenges related to population diversity in simpler image classification tasks, thereby inducing instability in the model.

\section{Performance Evaluation of QUSL in NISQ Environments}
\label{appendixD}

To validate QUSL's potential for practical application in the NISQ era, we conducted simulated experiments using the quantum noise library provided by Pennylane. We added composite noise with a degree of $0.045$ (including bit flip, phase flip, and depolarizing channel noise) to the experiments. Additionally, to improve efficiency, we compressed the images through downsampling. Five independent trials were conducted in both clean and noisy environments with CIFAR\_10:

\begin{table}[htb]
	\renewcommand{\arraystretch}{1.2}
	\caption{Performance of QUSL in noise-free and noisy environments}
	\centering
	\scriptsize
	\begin{threeparttable}
		\begin{tabularx}{\linewidth}{@{}X*{5}{>{\centering\arraybackslash}p{0.09\linewidth}}>{\centering\arraybackslash}p{0.2\linewidth}@{}}
			\toprule
			Environment & 1st & 2nd & 3rd & 4th & 5th & Average $(\pm std)$ \\
			\midrule
			Noise-free & 0.8094 & 0.7941 & 0.7824 & 0.7778 & 0.7755 & $0.7878(\pm0.0143)$ \\
			Noisy & 0.8063 & 0.7492 & 0.7246  & 0.7411 & 0.7067 & $0.7226(\pm0.0354)$ \\
			\bottomrule
		\end{tabularx}
	\end{threeparttable}
\end{table}

Results show that QUSL's performance remained stable in noisy environments, with only a slight decrease in robustness. We speculate that the noise primarily affected the consistency in the triplet construction process, thereby influencing feature extraction. Nevertheless, QUSL's excellent performance in noisy environments demonstrates its potential for application on actual NISQ devices.

\section{Performance Comparison of QUSL with Classical Methods}
\label{appendixE}

To evaluate QUSL's performance, we designed a set of comparative experiments in CIFAR$\_$10. The classical baseline used a medium-scale fully connected neural network with $4$ hidden layers and approximately $1$ million parameters, also employing the triplet strategy.

\begin{table}[htb]
	\renewcommand{\arraystretch}{1.2}
	\caption{Performance comparison of Classical Network, SliQ, and QUSL}
	\centering
	\scriptsize
	\begin{threeparttable}
		\begin{tabularx}{\linewidth}{@{}X*{5}{>{\centering\arraybackslash}p{0.09\linewidth}}>{\centering\arraybackslash}p{0.2\linewidth}@{}}
			\toprule
			Method & 1st & 2nd & 3rd & 4th & 5th & Average $(\pm std)$ \\
			\midrule
			Classical & 0.833 & 0.771 & 0.766 & 0.745 & 0.742 & $0.771(\pm0.036)$ \\
			SliQ & 0.743 & 0.742 & 0.742 & 0.686 & 0.619 & $0.706(\pm0.052)$ \\
			QUSL & 0.890 & 0.792 & 0.769 & 0.766 & 0.731 & 0.790 $(\pm$0.059) \\
			\bottomrule
		\end{tabularx}
	\end{threeparttable}
\end{table}

The experimental results show that the performance of the three networks is roughly on par, which is already an encouraging outcome. Notably, SliQ achieved comparable performance to the classical network using only about $200$ parameters, demonstrating the potential of quantum methods in parameter efficiency. QUSL outperforms SliQ in effectiveness, and also shows better performance in quantum circuit efficiency and quantum resource utilization, making it more adaptable to current quantum hardware conditions.

\end{document}